\documentclass[paper,notoc]{JHEP3}

\usepackage{epsfig}
\usepackage{amsmath}
\usepackage{amssymb}
\usepackage{graphicx}

\usepackage{verbatim}   % for \begin{comment} \end{comment}
\usepackage{multirow}   % for multirow's in tables

\allowdisplaybreaks[1]

\DeclareGraphicsExtensions{eps,ps}

\def \refeq#1{(\ref{#1})}  
\def \refsec#1{Section~\ref{#1}}

\def \refapp#1{Appendix~\ref{#1}}
\def \reffig#1{Figure~\ref{#1}}
\def \reftab#1{Table~\ref{#1}}

\def \nn{\nonumber\\}

\def \aver#1{\left\langle #1 \right\rangle}

\def \order#1{ {\cal O} \left( #1 \right) }

\def \re{\textrm{Re}}
\def \im{\textrm{Im}}

\def \cLdB1{{{\cal L}_{\Delta B = 1}^{\rm EW}}} % Delta_B = 1 effective Lagrangian
\def \BR{{\cal B}}                               % branching ratio 

\def \Op{{\cal O}}

\def \One{\leavevmode\hbox{\small1\kern-3.6pt\normalsize1}} % unit matrix

\def \MeV{{\rm \; MeV}}
\def \GeV{{\rm \; GeV}}

\def \alS{\alpha_s}        % strong coupling
\def \alE{\alpha_e}        % electro-magnetic coupling
\def \GF{G_F}              % Fermi coupling
\def \LamConf{{\Lambda_{}}}    %  confinement scale
\def \BRll#1#2#3{{\BR(#1) |_{[#2] #3}}}

\def \thl {{\theta_l}}
\def \thK {{\theta_{K^*}}}

\def \azeL{{A_0^L}}
\def \azeR{{A_0^R}}
\def \apaL{{A_\|^L}}
\def \apaR{{A_\|^R}}
\def \apeL{{A_\bot^L}}
\def \apeR{{A_\bot^R}}

\def \BXclv  {\bar{B} \to X_c l \bar\nu_l}
\def \BXsgamma {\bar{B} \to X_s \gamma}

\def \Bstomm{{\bar{B}_s \to \bar{\mu} \mu }}

\def \BXsll{\bar{B} \to X_s \bar{l} l}

\def \BtoKast{{ B \to K^{\ast}}}
\def \B0toK0ast{{ B^0 \to K^{\ast 0}}}
\def \BtoKll{{\bar{B} \to \bar K \bar{l} l}}

\def \BtoKastll{{\bar{B} \to \bar K^\ast \bar{l} l }}

\def \B0toK0mumu{{ \bar{B}^0 \to K^0 \bar{\mu} \mu}}
\def \BtoKKpill{{ \bar B \to \bar K^\ast (\to \bar K \pi) \bar{l} l}}

\def \barB0toKKpill{{ \bar{B}^0 \to \bar{K}^{\ast 0} (\to K^- \pi^+) \bar{l} l}}
\def \B0toKKpill{{ B^0 \to K^{\ast 0} (\to K^+ \pi^-) \bar{l} l}}

\def\MSbar{{\overline{\rm MS}}}    % MS-bar

\def\D0{D\O}  \def\d0{D\O}

\title{CP Asymmetries in $\BtoKKpill$ and \\
Untagged $\bar B_s,B_s \to \phi (\to K^+ K^-) \bar l l$ Decays at NLO}

\author{Christoph Bobeth, Gudrun Hiller and Giorgi Piranishvili \\
  Institut f{\"u}r Physik, Technische Universit{\"a}t Dortmund, D-44221
  Dortmund, Germany}
\date{\today}

\abstract{ 

The decay $\BtoKKpill$ offers great opportunities to explore the physics
at and above the electroweak scale by means of an angular analysis.
We investigate the physics potential of the seven CP asymmetries
plus the asymmetry in the rate, working at low dilepton mass using 
QCD factorization at next-to leading order (NLO).
The $b \to s$ CP asymmetries are doubly Cabibbo-suppressed $\lesssim 1\%$
in the Standard Model and its extensions where the CKM matrix is 
the only source of CP violation.
Three CP asymmetries are T-odd, and can be ${\cal{O}}(1)$ in the presence
of non-standard CP violation. The T-even asymmetries can reach 
${\cal{O}}(0.1)$, limited by the small strong phases
in the large recoil region. We furthermore point out an easy way to 
measure CP phases from time-integrated, untagged
$\bar B_d ,B_d \to K^* (\to K^0 \pi^0) \bar l l$  and
$\bar B_s,B_s \to \phi (\to K^+ K^-) \bar l l$ decays.
Analyses of these CP asymmetries can rule out, 
or further support the minimal description
of CP violation through the CKM mechanism.  
Experimental studies are promising for
(super) flavor factories and at hadron colliders.
}


\preprint{DO-TH 08/03}

\begin{document}

%--------+---------+---------+---------+---------+---------+---------+---------+
\section{Introduction}

The quest for physics beyond the Standard Model (SM) is of highest priority 
at current and future flavor facilities \cite{superbLHCb}. 
A promising direction is to look for CP symmetry breaking effects that 
cannot be accounted for with the SM`s very predictive CKM-mechanism of CP and 
flavor violation for the quarks.
Rare $b \to s $ transitions are such sensitive probes, since
all CP violating effects in the SM receive a universal suppression
of order $10^{-2}$ from the CKM matrix elements. Ongoing  
experimental efforts include those with hadronic decays such as
$\bar B_s, B_s \to J/\Psi \phi$ at the Tevatron
\cite{tevatronBs} and $b \to s \bar s s$ induced penguin modes, 
see, e.g., \cite{Barberio:2006bi} for recent data.

We focus here on exclusive semileptonic decays 
induced by $b \to s \bar l l$, $l=e, \mu$, to explore
the borders of the SM. The decays  $\BtoKll$ and  $\BtoKastll$ have already 
been observed, with their rate being in agreement with the SM.
Investigations of more involved observables such as dilepton mass spectra,
lepton angle distributions and dimuon to dielectron ratios are currently
being pursued \cite{Abe:2004ir,Ishikawa:2006fh,Aubert:2006vb}. 
The decay $\BtoKastll$ with subsequent $\bar K^* \to \bar K \pi$ allows
to extract further physics information through an angular analysis of the 
$\bar K \pi \bar l l$ final state \cite{Kruger:1999xa}, especially on 
the chirality content \cite{RHcurrents, Kruger:2005ep,Lunghi:2006hc}.

Here we study the eight CP asymmetries in $\BtoKKpill$ decays, the one in the
decay rate plus seven more requiring angular information.
We calculate the matrix elements to next-to-leading order 
(NLO) in the strong coupling and to lowest order (LO)
in the $1/E$ expansion \cite{Charles:1998dr}, where $E$ denotes 
the energy of the emitted $\bar K^*$ in $\bar B$ rest frame.
We work within QCD factorization (QCDF), which has been applied to
$\BtoKastll$ decays \cite{Beneke:2001at, Beneke:2004dp}, for
analyses in soft-collinear effective theory (SCET), see \cite{Ali:2006ew}.
Previous works on CP asymmetries in the angular distributions 
\cite{Kruger:1999xa,FK2003,Kim:2007fx} employed naive 
factorization, see \cite{Chen:2002bq} for a PQCD study.

By means of the large energy limit, the QCDF framework used in our analysis 
holds for small dilepton invariant masses, and predicts small strong phases:
Lowest order quark loops are either close to or below threshold, 
hence real (charm quarks), or CKM suppressed or induced by small 
penguin contributions (the lighter quarks).
Other sources of strong phases are from subleading 
spectator effects or higher order in $\alS$.

This highlights T-odd CP asymmetries, three of which
are accessible with the angular analysis.
(The T-transformation reverses the sign of all particle momenta and spins.)
The important feature is that the T-odd asymmetries are 
$\propto \cos \Delta_S \sin \Delta_W$, where $\Delta_S$ and $\Delta_W$
denote the differences of strong and weak phases, respectively.
While the T-even CP asymmetries $\propto \sin \Delta_S \sin \Delta_W$ vanish
for small strong phases, the T-odd asymmetries 
exhibit maximal sensitivity to CP violation in this limit.

Time-dependent CP asymmetries in
$\bar B_d ,B_d \to K^* (\to K^0 \pi^0) \bar l l$ 
have also been considered in \cite{Kim:2007fx}. These measurements require 
large amounts of data. We suggest here to use the CP-odd property of four 
of the asymmetries to extract them from an untagged and time-integrated data 
set. It has been known that such data sets are useful to access 
CP violation in angular distributions, e.g., 
\cite{Fleischer:1996aj,Sinha:1996sv}. We work out the corresponding 
CP-sensitive observables in 
$\bar B_d ,B_d \to K^* (\to K^0 \pi^0) \bar l l$ 
and $\bar B_s, B_s \to \phi (\to K^+ K^-) \bar l l$ decays.
The $K^+ K^- \bar l l$ final state is the same as the one of the
$\bar B_s, B_s \to \phi (\to K^+ K^-) J/\Psi (\to \bar l l)$ decays, 
which are already under experimental study including angular analysis 
\cite{tevatronBs}.

In \refsec{sec:diff:dist} we review the $\BtoKKpill$ angular distributions. 
CP asymmetries and possibilities of their measurement from 
double- and single-differential distributions are given in \refsec{sec:CP:asy}.
Prospects for $\bar B_d ,B_d \to K^* (\to K^0 \pi^0) \bar l l$ and
$\bar B_s, B_s \to \phi (\to K^+ K^-) \bar l l$ decays without tagging
are examined in \refsec{sec:mixing}. SM predictions and theoretical 
uncertainties are presented in \refsec{sec:SM:num}. In \refsec{sec:NP} 
we model-independently investigate the impact of New Physics (NP) on the 
CP asymmetries. To do so, we work out constraints from other rare decay data, 
$\bar B \to X_s \bar ll, X_s\gamma$, 
the forward-backward asymmetry in $\bar B \to \bar K^* \bar l l$ and 
time-dependent CP asymmetries in 
$\bar B_d, B_d \to K^* (\to K^0 \pi^0) \gamma$. We summarize in 
\refsec{sec:conclusions}. Various appendices \ref{app:Ii}-\ref{app:num:CP:asy} 
contain details of the calculation of the CP asymmetries.

%--------+---------+---------+---------+---------+---------+---------+---------+
\section{Full Angular Distribution 
\label{sec:diff:dist}}

In this section we review the angular distribution of the exclusive
decay $\BtoKKpill$ and its conjugate decay. 
Throughout this work we use $\bar B \equiv (b \bar q)$ for $q=u,d$, 
$\bar B_s \equiv (b \bar s)$ and $\bar K, \bar{K}^{*} \equiv (s \bar q)$.
We illustrate the kinematics for neutral mesons decaying to charged
particles. Charged $B$-decays can be treated 
analogously. We discuss meson mixing effects and 
$\bar B_s \to \phi (\to K^+ K^-) \bar l l$ decays in \refsec{sec:mixing}.

The full angular distribution of the decay $\barB0toKKpill$
can be written in the limit of an on-shell decaying $K^\ast$ resonance 
as a 4-differential distribution  \cite{Kruger:1999xa,Kruger:2005ep} 
\begin{equation}
  \label{eq:d4Gam}
  \frac{d^4\Gamma}{dq^2\, d\cos\thl\, d\cos\thK\, d\phi} =
   \frac{3}{8\pi} J(q^2, \thl, \thK, \phi),
\end{equation}
where the lepton spins have been summed over.
Here, $q^2$ is the dilepton invariant mass squared, that is, 
$q^\mu$ is the sum of $p_{l^+}^\mu$ and $p_{l^-}^\mu$, the four momenta of the
positively and negatively charged lepton, respectively.
Furthermore, $\thl$ is defined as the angle between the negatively charged 
lepton and the $\bar{B}$ in the dilepton center of mass system (c.m.s.) and 
$\thK$ is the angle between the Kaon and the $\bar{B}$ 
in the $(K^-\pi^+)$ c.m.s.. We denote by  $\mathbf{p}_i$ the three momentum 
vector of particle $i$ in the $\bar{B}$ rest frame. Then, $\phi$ is given by 
the angle between $\mathbf{p}_{K^-} \times \mathbf{p}_{\pi^+}$ and
$\mathbf{p}_{l^-} \times \mathbf{p}_{l^+}$, i.e., the angle between the 
normals of the $(K^-\pi^+)$ and $(l^-l^+)$ planes.
The full kinematically accessible phase space is bounded by
\begin{align}
  4 m_l^2 & \leqslant q^2\leqslant (M_B - M_{K^*})^2, &
  -1 & \leqslant \cos\thl \leqslant 1, &
  -1 & \leqslant \cos\thK \leqslant 1, &
   0 & \leqslant \phi     \leqslant 2 \pi, 
\end{align}
where $m_l, M_B$ and $M_{K^*}$ denote the mass of the lepton, $B$ meson 
and the $K^*$, respectively. For an off-resonance 
$\bar B \to \bar K \pi \bar l l$ study, see \cite{Grinstein:2005ud}.

The dependence of the decay distribution \refeq{eq:d4Gam}
on the angles $\thl,\, \thK$ and $\phi$ can be made explicit as  
\begin{align}
  J(q^2, \thl, \thK, \phi)& = J_1^s \sin^2\thK + J_1^c \cos^2\thK
      + (J_2^s \sin^2\thK + J_2^c \cos^2\thK) \cos 2\thl
\nonumber \\       
    & + J_3 \sin^2\thK \sin^2\thl \cos 2\phi 
      + J_4 \sin 2\thK \sin 2\thl \cos\phi 
      + J_5 \sin 2\thK \sin\thl \cos\phi
\nonumber \\      
    & + J_6 \sin^2\thK \cos\thl 
      + J_7 \sin 2\thK \sin\thl \sin\phi
\nonumber \\ 
    & + J_8 \sin 2\thK \sin 2\thl \sin\phi
      + J_9 \sin^2\thK \sin^2\thl \sin 2\phi , 
  \label{eq:I:func}
\end{align}
where the coefficients $J_i^{(a)} =J_i^{(a)}(q^2)$ for $i = 1, \ldots, 9$ 
and $a = s,c$ are functions of the dilepton mass. 
In the following we suppress the $q^2$-dependence for brevity also in 
expressions derived from the $J_i^{(a)}$.

The angular coefficients $J_i^{(a)}$ can be expressed through the 
$K^*$ transversity amplitudes
$A_i(q^2)$ with $i = \{\perp, \parallel, 0, t\}$, see \refapp{app:Ii}.
Note that not all the  $J_i^{(a)}$ are independent, for example, 
for vanishing lepton masses
\begin{equation}
  3 J_1^s = J_2^s , ~~~~~~~~~~~~~ J_1^c =- J_2^c . 
\end{equation}
Furthermore, in the absence of right-handed currents  $J_3=J_9=0$ up to 
power corrections.

The corresponding distribution of the CP conjugated decay $\B0toKKpill$ can 
be written as
\begin{equation}
  \label{eq:d4barGam}
  \frac{d^4\bar{\Gamma}}{dq^2\, d\cos\thl\, d\cos\thK\, d\phi} =
     \frac{3}{8\pi} \bar{J}(q^2, \thl, \thK, \phi).
\end{equation}
Here, $\theta_{K^*}$ denotes the angle between the Kaon and the $B$ meson
in the $(K^+\pi^-)$ c.m.s.. The definiton of $\theta_l$ is identical for both
$B$ and $\bar B$ decays. The angle $\phi$ for $B$ decays is given by 
the angle between $\mathbf{p}_{K^+} \times \mathbf{p}_{\pi^-}$ and
$\mathbf{p}_{l^-} \times \mathbf{p}_{l^+}$.
Therefore, in the limit of unbroken CP, the distributions for
$B$ and $\bar B$ mesons are equal
under the combined transformations 
$\theta_l \to \theta_l-\pi$ and $\phi \to -\phi$.
The function $\bar{J}$ is hence obtained from $J$ 
in \refeq{eq:I:func} by the replacements 
\begin{align}
  \label{eq:CP:I}
  J_{1,2,3,4,7}^{(a)} & \to  \bar{J}_{1,2,3,4,7}^{(a)}, &
  J_{5,6,8,9} & \to -\bar{J}_{5,6,8,9} ,
\end{align}
where $\bar{J}_i^{(a)}$ equals $J_i^{(a)}$ with all
weak phases being conjugated \cite{Kruger:1999xa}. 

With its rich multi-dimensional structure, the angular distributions 
in \refeq{eq:d4Gam} and \refeq{eq:d4barGam} have  sensitivity to various 
effects modifying the
SM, such as CP violation beyond CKM and/or right-handed currents.
Given sufficient data, all $J_i^{(a)}$ and  $\bar J_i^{(a)}$ can 
in principle be completely measured from the full angular distribution
in all three angles $\thl,\, \thK$  and $\phi$.

The familiar dilepton invariant mass spectrum for 
$\bar B \to \bar K^* \bar l l$ decays
can be recovered after integration over all angles as
\begin{equation}
  \label{eq:rate}
  \frac{d\Gamma}{dq^2} = J_1 - \frac{J_2}{3},  ~~~ \mbox{where} ~~J_{1,2} \equiv 2 J^s_{1,2} + J^c_{1,2} .
\end{equation}
The (normalized) forward-backward asymmetry $A_{\rm FB}$ 
is given after full $\phi $ and $\thK$ integration as
\footnote{Since we define the lepton angle $\thl$ with respect to the $l^-$, 
our definiton of the forward-backward
asymmetry \refeq{eq:AFB:thl} differs from the one in other works using the 
$l^+$, e.g., \cite{Ishikawa:2006fh,Aubert:2006vb,Beneke:2001at, Beneke:2004dp,Ali:2002jg},
by a global sign.}

\begin{align}
  \label{eq:AFB:thl}
  A_{\rm FB}(q^2) & \equiv
     \left[\int_{0}^1 - \int_{-1}^0 \right] d\cos\theta_l\, 
          \frac{d^2\Gamma}{dq^2 \, d\cos\theta_l} \Bigg/\frac{d\Gamma}{dq^2}
   =  J_6 \Bigg/ \frac{d\Gamma}{dq^2}. 
\end{align}
By $d \bar \Gamma /dq^2$ and $\bar A_{\rm FB}(q^2)$ we refer to the 
corresponding spectra of the CP conjugated decays.

%
%--------+---------+---------+---------+---------+---------+---------+---------+
\section{CP Asymmetries \label{sec:CP:asy}}

CP violating effects in the angular distribution are signaled by 
non-vanishing differences between the ($q^2$-dependent) angular coefficients
\begin{equation}
  \Delta J_i^{(a)} =\Delta J_i^{(a)}(q^2) \equiv J_i^{(a)} - \bar{J}_i^{(a)}.
\end{equation} 

Of particular importance are the asymmetries related to
the coefficients $J_{7,8,9}$. These are odd under $\phi \to -\phi$, and hence 
induce T-odd asymmetries $\Delta J_{7,8,9}$ which are not suppressed by
small strong phases as predicted from QCDF.

The CP asymmetry in the dilepton mass distribution is commonly defined as, see
\refeq{eq:rate},
\begin{equation}
  \label{eq:diff:CPasy:1}
  A_{ \rm CP}(q^2) \equiv 
    \frac{d (\Gamma - \bar{\Gamma})}{dq^2} \Bigg/ \frac{d (\Gamma + \bar{\Gamma})}{dq^2}
  = \frac{1}{N_\Gamma} \left[ \Delta J_1 - \frac{\Delta J_2}{3}\right]
, ~~ N_\Gamma=N_\Gamma(q^2)= \frac{d (\Gamma + \bar{\Gamma})}{dq^2} .
\end{equation}
Following \cite{Kruger:1999xa}, we define in addition to $A_{\rm CP}$ 
seven normalized CP asymmetries as
\begin{align}
  \label{eq:diff:CPasy:2}
  A_i(q^2) &  \equiv \frac{2 \Delta J_i}{N_\Gamma}  ~~\mbox{for}~ i  = 3,6,9, &
  A_i^D(q^2) & \equiv - \frac{2 \Delta J_i}{N_\Gamma}  ~~\mbox{for}~ i  = 4,5,7,8. 
\end{align}
Note that up to differences in the normalization $A_6$ equals the 
forward-backward CP asymmetry $A_{\rm FB}^{\rm CP}$ advocated to search for 
non-standard CP violation in the decay $\BtoKastll$ 
\cite{Buchalla:2000sk,Kruger:2000zg}, see \refeq{eq:AFB:thl}.
For $q^2$-integrated quantities we introduce the notation
\begin{equation}
  \aver{X} = \int_{q^2_{\rm min}}^{q^2_{\rm max}} dq^2 \, X(q^2),
\end{equation}
where the integration from $q^2_{\rm min}$ to $q^2_{\rm max}$ should be in 
the low dilepton mass region in order to use
the $1/E$ expansion of QCD for theory predictions, see \refapp{sec:transamp}.
We then define the normalized $q^2$-integrated CP asymmetries as
\begin{align} \label{eq:averageAi}
\aver{A_i} & \equiv 2 \frac{\aver{\Delta J_i}}{\aver{N_\Gamma}}  ~~\mbox{for}~
   i  = 3,6,9, &
\aver{A_i^D} & \equiv  - 2 \frac{\aver{\Delta J_i}}{\aver{N_\Gamma}} 
~~\mbox{for}~   i  = 4,5,7,8,
\end{align}
where the numerator and the denominator are 
integrated with the same $q^2$ cuts.

The CP asymmetries $\aver{A_i}$ ($i=3,6,9$) can, for example, be 
extracted from the double-differential distribution in $\theta_l$ and $\phi$,
\begin{align}
  \frac{d^2 \aver{\Gamma}}{d\cos\thl\, d\phi} & = \frac{1}{4 \pi} \big\{ 
     \aver{J_1} + \aver{J_2} \cos 2\thl + 2 \aver{J_3} \sin^2\thl \cos 2\phi 
\nonumber\\
  & \hspace{1cm}  
    + 2 \aver{J_6} \cos\thl + 2 \aver{J_9} \sin^2\thl \sin 2\phi \big\} ,
  \label{eq:d2G:thl:phi}
\end{align}
which is obtained from integrating \refeq{eq:d4Gam} over  $\thK$.
After full $\thl$-integration follows
\begin{align}
  \frac{d \aver{\Gamma}}{d\phi} & = \frac{1}{2 \pi} \left\{ 
     \aver{J_1} - \frac{\aver{J_2}}{3}
   + \frac{4}{3} \aver{J_3} \cos 2\phi + \frac{4}{3} \aver{J_9} \sin 2\phi \right\},
\end{align}
showing that $\aver{\Delta J_9}$ can be found from 
$d \aver{\Gamma + \bar{\Gamma}}/d\phi$,  whereas $\aver{\Delta J_3}$ can be 
obtained from $d \aver{\Gamma - \bar{\Gamma}}/d\phi$,
with $\aver{\Delta J_1} -\aver{\Delta J_2}/3$ from $A_{\rm CP}$ 
without angular study, see \refeq{eq:diff:CPasy:1}.

The measurement of the CP asymmetries $\langle A_i^{D} \rangle$ ($i= 4,5,7,8$)
requires binning into $\cos\theta_{K^*}$ as
\begin{align}
    \label{eq:d2Afb:thk}
    \frac{d^2 \aver{A_\thK}}{d\cos\thl\, d\phi} & \equiv
      \left[\int_{0}^1 - \int_{-1}^0 \right] d\cos\thK\,
      \frac{d^3 \aver{\Gamma}}{d\cos\thK\, d\cos\thl\, d\phi}
  \\  
    & = \frac{1}{2\pi} \big\{ 
       \aver{J_4} \sin 2\thl \cos\phi + \aver{J_5} \sin\thl \cos\phi
     + \aver{J_7} \sin\thl \sin\phi  + \aver{J_8}\sin 2\thl \sin\phi \big\}.
  \nonumber
\end{align}
{}From here follows upon full $\thl$-integration
\begin{align}
    \frac{d \aver{A_\thK}}{d\phi} & = 
     \frac{1}{4} \left\{ \aver{J_5} \cos\phi + \aver{J_7} \sin\phi \right\}.
\end{align}
We learn that $\aver{\Delta J_5}$ can be extracted from
$d \aver{A_\thK + \bar{A}_\thK}/d\phi$ 
  whereas $\aver{\Delta J_7}$ can be obtained from 
$d \aver{A_\thK - \bar{A}_\thK}/d\phi$.
The double asymmetry in $\thK$  and $\thl$,
\begin{align}
  \frac{d \aver{A_{\thK,\thl}}}{d\phi} & \equiv
  \left[\int_{0}^1 - \int_{-1}^0 \right] d\cos\thl\, \frac{d^2 \aver{A_\thK}}{d\cos\thl\, d\phi}
 = \frac{2}{3 \pi} \left\{ \aver{J_4} \cos\phi + \aver{J_8} \sin\phi \right\},
\end{align}
allows to obtain $\aver{\Delta J_4}$ from 
$d \aver{A_{\thK,\thl} -\bar{A}_{\thK,\thl}}/d\phi$, 
whereas $\aver{\Delta J_8}$ can be extracted 
from $d \aver{A_{\thK,\thl} + \bar{A}_{\thK,\thl}}/d\phi$.

The latter considerations demonstrate how the CP violating angular 
coefficients $\Delta J_i$ for $i=3,4,5,7,8,9$ can
be extracted from distributions in the angle $\phi$. The quantity
$\Delta J_6$ can be measured easiest from the $\cos \thl$-distribution, 
i.e., by adding the (numerators of the) forward-backward asymmetries
 $A_{\rm FB}$ and $\bar{A}_{\rm FB}$, see \refeq{eq:AFB:thl}.
Note that only $A_3,A_6$ and $A_9$ can be obtained from a genuinely 
single-differential distribution. $A_9$ is the only T-odd asymmetry 
with this property.

Analytical expressions for all CP asymmetries at NLO in terms of the
short distance coefficients from the electroweak Hamiltonian in 
\refapp{sec:eff:Ham} are presented in \refapp{app:CP:asy}.
In particular, we include NLO $\alpha_s$-corrections 
thus present in this work the first analyses of CP asymmetries in the 
$\BtoKKpill$ angular distributions at this order.

Corrections from $B_d-\bar B_d$ mixing to the time integrated CP 
asymmetries in flavor specific (self-tagging) final states are
of the order $|A_{\rm SL}^d | \lesssim {\cal{O}}(10^{-3})$  
\cite{Yao:2006px} and can be neglected. Here, $A_{\rm SL}^d$ denotes the 
semileptonic asymmetry into wrong sign leptons in the $B_d$-system.

%
%--------+---------+---------+---------+---------+---------+---------+---------+
\section{CP Asymmetries without Tagging \label{sec:mixing}}

With the $J_i$, $i = 5,6,8,9$ being odd under CP, see \refeq{eq:CP:I}, 
the corresponding CP asymmetries  can be extracted from 
$d \Gamma + d \bar \Gamma$, i.e., without identifying the flavor of the 
initial $b$ quark. This feature is very useful for   
$\bar B_d ,B_d \to K^* (\to K^0 \pi^0) \bar l l$ and
$\bar B_s, B_s \to \phi (\to K^+ K^-) \bar l l$ decays, which unlike
 $\bar B_d,B_d \to K^* ( \to K^\mp \pi^\pm) \bar l l$ or charged 
$B$-decays, are not self-tagging. We focus here on $B_s$-decays to 
CP eigenstates, but the formalism equally applies to 
$B_d$-decays after the corresponding replacements.

Both $\bar B_s$ and $B_s$  angular distributions are described by the angles
$\theta_l$, $\thK$ and $\phi$. These are defined in complete analogy 
with $\BtoKKpill$ decays, see \refsec{sec:diff:dist}:
$\thl$ is the angle between the negatively charged lepton and the 
$\bar{B}_s/B_s$ in the dilepton c.m.s., $\thK$ denotes the angle between 
the $K^-$ and the $\bar{B}_s/B_s$ in the $(K^-K^+)$ c.m.s. and $\phi$ is  
the angle between $\mathbf{p}_{K^-} \times \mathbf{p}_{K^+}$ and
$\mathbf{p}_{l^-} \times \mathbf{p}_{l^+}$.

To account for mixing, time-dependent transversity amplitudes
need to be introduced:
\begin{eqnarray}
A_a(t) \equiv A( \bar B_s(t) \to \phi (\to K^+ K^-)_a \bar l l)  ,~~~
\bar A_a(t) \equiv A( B_s(t) \to \phi (\to K^+ K^-)_a \bar l l) ,
\end{eqnarray}
where $A_a(t), ( \bar A_a(t))$ denotes the amplitude for a meson born 
at time $t=0$ as a $\bar B_s$, ($B_s$) decaying through the 
transversity amplitude 
$a=\perp,\parallel,0$ at later times $t$. 
Here we use for brevity $A_a(t)$ for both $A_a^{L}(t)$ and $A_a^R(t)$.
The formulae for the unmixed transversity amplitudes, i.e., the ones
at $t=0$ can be taken from \refapp{sec:transamp} with the requisite 
replacements in masses and hadronic parameters and differences
in the spectator effects given in \cite{Beneke:2004dp} to account for the 
$\bar B_s \to \phi$ transitions.

Untagged rates $d \Gamma + d \bar \Gamma$ can then be written as
($a,b=\perp,\parallel,0$) \cite{Fleischer:1996aj}
\begin{eqnarray}
\bar A_{b}^*(t) \bar A_a(t)+A_b^*(t) A_a(t)=
\frac{1}{2}
A( B_s \to \phi (\to K^+ K^-)_b \bar l l)^* 
A( B_s\to \phi (\to K^+ K^-)_a \bar l l) 
\nonumber\\
\times\left[\left(1+\eta_a \eta_b  \xi_b^* \xi_a \right)
\left(e^{-\Gamma_L t}+e^{-\Gamma_H t}\right)+\left( \eta_b \xi_b^*
+\eta_a \xi_a \right) \left(e^{-\Gamma_L t}-e^{-\Gamma_H t}\right)\right].
\label{eq:untagged}
\end{eqnarray}
Here,  $\eta_{\parallel,0} =+1$ and $\eta_\perp =-1$ are the CP eigenvalues 
of the final state and
$\Gamma_{L (H)}$ denotes the width of the lighter (heavier) mass eigenstate.
We neglect CP violation in mixing, which is bounded by the semileptonic 
asymmetry in the $B_s$-system 
$|A_{\rm SL}^s| \lesssim {\cal{O}}(10^{-2})$ \cite{tevatronBs}.
Furthermore, 
\begin{equation} \label{eq:xia}
\xi_a = e^{-i \Phi_M} \frac{A( \bar B_s \to \phi (\to K^+ K^-)_a \bar l l) }{A( \bar B_s \to \phi (\to K^+ K^-)_a \bar l l) (\delta_W \to - \delta_W)} ,
\end{equation}
where $(\delta_W \to - \delta_W)$ 
implies the conjugation of all weak phases in the denominator and 
$\Phi_M$ denotes the phase of the $B_s - \bar B_s$ mixing.
It is very small in the SM, $\Phi_M^{\rm SM}= 2 \arg(V_{ts}^* V_{tb})$.

The CP asymmetries $\Delta J_i(t)=J_i(t) -\bar J_i(t)$,  $i=5,6,8,9$
are then obtained by taking the real and imaginary 
parts of \refeq{eq:untagged}, adding or subtracting 
$\bar A_{b}^{k*}(t) \bar A^k_a(t)+A^{k*}_b(t) A^k_a(t)$ for
$k=L$ and $k=R$, and taking into account normalization factors
depending on the angular coefficient $J_i$,
see \refapp{app:Ii}. After time-integration follows from  \refeq{eq:untagged}
\begin{align}
\int _0^\infty dt(\bar A_{b}^*(t) \bar A_a(t)+A_b^*(t) A_a(t)) &=
A( B_s \to \phi (\to K^+ K^-)_b \bar l l)^* 
A( B_s\to \phi (\to K^+ K^-)_a \bar l l) 
\nonumber\\
\times \frac{1}{\Gamma (1-y^2)} & 
\left[\left(1+\eta_a \eta_b  \xi_b^* \xi_a \right)
-y \left( \eta_b \xi_b^*+\eta_a \xi_a \right) \right] , ~~~~ y=\frac{\Delta \Gamma}{2 \Gamma} ,
\label{eq:untagged-tint}
\end{align}
where $\Gamma =(\Gamma_L +\Gamma_H)/2$ and the width difference
$\Delta \Gamma =\Gamma_L -\Gamma_H$.

This expression \refeq{eq:untagged-tint} becomes transparent if one
neglects strong phases, where
$\xi_a=e^{-i (\Phi_M -2 \Phi_a)}$,
$ \Phi_a \equiv \arg(A( \bar B_s \to \phi (\to K^+ K^-)_a \bar l l)$.
For $a \neq b$, $\eta_a =-\eta_b$ we obtain
\begin{align}
\int _0^\infty dt \, \im(\bar A_{b}^*(t) \bar A_a(t)+A_b^*(t) A_a(t))
 =  \frac{2}{\Gamma (1-y^2)} |A( B_s \to \phi (\to K^+ K^-)_b \bar l l)| 
\nonumber \\ \label{eq:imaa}
\times |A( B_s\to \phi (\to K^+ K^-)_a \bar l l) | \cdot
\left[  \sin (\Phi_a-\Phi_b) - y \eta_a \sin (\Phi_M -\Phi_a -\Phi_b) \right]
, \\
\int _0^\infty dt \sum_a( |\bar A_a(t)|^2 + |A_a(t)|^2) =
\frac{2}{\Gamma (1-y^2)} \sum_a |A( B_s \to \phi (\to K^+ K^-)_a \bar l l)|^2 
\nonumber \\ \label{eq:absaa}
\times \left[ 1 - y \eta_a \cos (\Phi_M -2 \Phi_a) \right] ,
\end{align}
where \refeq{eq:imaa} gives the asymmetries related to 
$J_{8,9}$ and \refeq{eq:absaa} is needed for normalization, that is, 
gives (twice) the  CP averaged decay rate. It also exhibits  
sensitivity to CP phases. The T-even asymmetries associated with $J_{5,6}$ 
vanish with no strong phases present.

We define ($q^2$-dependent) CP-odd CP asymmetries $A_i^{(D)mix}$ as
($a=\perp,\parallel,0$, $k=L,R$)
\begin{align} \nonumber
A_i^{mix}(q^2) & \equiv 2 \frac{ \int_0^\infty dt \Delta J_i(t)}{ \int_0^\infty dt \sum_{a,k}( |\bar A^k_a(t)|^2 + |A^k_a(t)|^2)}  ~~\mbox{for}~~
   i  = 6,9, \\
A_i^{D mix}(q^2) & \equiv - 2 \frac{ \int_0^\infty dt \Delta J_i(t)}{ \int_0^\infty dt \sum_{a,k}( |\bar A^k_a(t)|^2 + |A^k_a(t)|^2)} ~~\mbox{for}~~ i  = 5,8,
\label{eq:Aimix}
\end{align}
which match the CP asymmetries of the flavor-specific, unmixed decays 
\refeq{eq:diff:CPasy:2} for $y \to 0$. 

For $B_d$-mesons, $y$ is below $10^{-2}$ \cite{Yao:2006px}, and the untagged 
and time-integrated $K^0 \pi^0$ final states yield the same information
on CP violation as the ones with $K^\mp \pi^\pm$ discussed in 
\refsec{sec:CP:asy}, or charged $B$-decays.
(For early works with $y=0$, see \cite{Sinha:1996sv}).
For $B_s$-mesons the width difference is larger, 
$y \sim {\cal{O}}(0.1)$ \cite{Yao:2006px}, and interference effects become 
observable with the mixing phase $\Phi_M$. The latter is currently
under intense experimental study and only poorly determined to date,
see, e.g., \cite{tevatronBs}.

We refrain in this work from presenting a dedicated numerical 
analysis for the $\bar B_s, B_s \to \phi (\to K^+ K^-) \bar l l$ 
observables $A_i^{(D)mix}$: The presumably dominant part independent of 
the width difference can be inferred from $\BtoKKpill$ decays by $SU(3)$. 
The biggest corrections such as those from the form factors and 
phase space are expected to cancel in the asymmetries. On the other hand,
discrepancies in the 
CP asymmetries between $B_d$- and $B_s$-processes at ${\cal{O}}(y)$ can 
be attributed to the mixing parameters $y$ and $\Phi_M$.

%
%--------+---------+---------+---------+---------+---------+---------+---------+
\section{Standard Model Predictions \label{sec:SM:num}}

CP asymmetries in the decays of hadrons are in the SM  solely induced by the 
CKM matrix. For the $b \to s$ transitions under consideration here, 
the requisite weak phase difference stems from
$\hat{\lambda}_u = V_{ub}^{} V_{us}^\ast/V_{tb}^{} V_{ts}^\ast$.
Therefore, all CP asymmetries in $\BtoKKpill$ decays 
discussed here receive an overall suppression by
$\im [\hat{\lambda}_u] \simeq \bar \eta \lambda^2$ of order $10^{-2}$, 
where $\lambda $ and $\bar \eta$ denote 
parameters of the Wolfenstein parametrization of the CKM matrix.

\begin{table}
%\TABLE[ht]{
\begin{center}
\begin{tabular}{||l|l||}
\hline \hline
  $\lambda = 0.2258^{+0.0016}_{-0.0017}$ \hfill $(95\%\, {\rm C.L.})$ \cite{CKMfitter:homepage} 
& 
  $\BR(\BXclv)=(10.57 \pm 0.15) \%$  \hfill\cite{Yao:2006px} 
\\
  $|V_{cb}| = 0.0417 \pm 0.0013$ \hfill $(95\%\, {\rm C.L.})$ \cite{CKMfitter:homepage} 
& 
  $\tau_{B^0} = (1.530 \pm 0.009) \,  {\rm ps} $ \hfill\cite{Yao:2006px}  
\\
  $\bar{\rho} = [0.108,\, 0.243]$ \hfill $(95\%\, {\rm C.L.})$ \cite{CKMfitter:homepage}
&
  $\tau_{B_s} = (1.425 \pm 0.041) \,  {\rm ps} $ \hfill\cite{Yao:2006px}  
\\
  $\bar{\eta} = [0.288,\, 0.375]$ \hfill $(95\%\, {\rm C.L.})$ \cite{CKMfitter:homepage}
& 
  $f_{B_{u,d}} = (200 \pm 30) \MeV$ \hfill 
\\
  $\alS(m_Z) = 0.1176 \pm 0.0020$ \hfill \cite{Yao:2006px}
&
  $f_{B_{s}} = (240 \pm 30) \MeV$ \hfill \cite{Onogi:2006km}
\\
  $\alE(m_b) = 1/133$
&
  $\lambda_{B,+}(1.5 \GeV) = (0.458 \pm 0.115) \GeV$ \hfill \cite{Beneke:2004dp, Braun:2003wx}
\\
  $m_W = 80.403  \GeV$ \hfill \cite{Yao:2006px}
&
  $f^{K^*}_{\perp}(1 \GeV) = (185 \pm 10) \MeV $ \hfill \cite{Ball:2005vx}
\\
  $m_t^{pole} = (170.9 \pm 1.8) \GeV$ \hfill \cite{:2007bx}
&
  $f^{K^*}_{\parallel} = (217 \pm 5) \MeV $ \hfill \cite{Yao:2006px}
\\
  $m_b=(4.6 \pm 0.1) \GeV$ \hfill \cite{Beneke:2001at}
& 
  $a^{\perp,\parallel}_{1,K^*}(1 \GeV) = 0.1 \pm 0.07$ \hfill \cite{Ball:2004rg}
\\
  $m_c^{pole}=(1.4 \pm 0.2) \GeV$
& 
  $a^{\perp,\parallel}_{2,K^*}(1 \GeV) = 0.1 \pm 0.1$ \hfill \cite{Ball:2004rg}  
\\
\hline \hline
\end{tabular}
\end{center}
\caption{ \label{tab:num:input} The numerical input used in our analysis.
  We denote by $m_b$ the PS mass at the factorization scale $\mu_f=2 \GeV$.
  We neglect the strange quark mass throughout this work unless otherwise
stated. 
  The numerical input for the form factors $\xi_{\perp, \parallel}$
  is given in \refapp{app:formf}. }
\end{table}

We work out the SM CP asymmetries in $\BtoKKpill$ decays in the 
low-$q^2$ region using QCDF at NLO in $\alS$ and leading order $1/E$.
Analytical expressions for the asymmetries are given in \refapp{app:CP:asy}.
The CP asymmetries in the SM can be obtained by setting the NP Wilson 
coefficients $C_{7,9,10}^{('), {\rm NP}}= 0$, see \refapp{sec:eff:Ham} for the
effective Hamiltonian used. Details on the QCDF framework and the transversity 
amplitudes are given in \refapp{sec:transamp}. We take the $B \to K^*$ 
form factors from light cone QCD sum rules (LCSR) calculations  
\cite{Ball:2004rg}, see \refapp{app:formf}. 
Our numerical input is compiled in \reftab{tab:num:input}. 
We checked that our findings for the branching ratio and 
the forward-backward asymmetry of $\BtoKastll$ decays agree  
for the given input with \cite{Beneke:2001at, Beneke:2004dp}.
Our predictions always refer to neutral $B$-decays unless otherwise stated.

The three main uncertainties in the asymmetries come from the form factors
$\xi_\parallel$ and  $\xi_\perp$, the variation of the renormalization 
scale $\mu_b$ and the CKM parameters. We vary  the scale between $m_b/2$ 
and $2 m_b$ and allow for an uncertainty of $11\%$ and $14\%$ for 
$\xi_\perp$ and $\xi_\parallel$, respectively. The CKM input is given in 
\reftab{tab:num:input}. For the total uncertainty estimate, all three 
sources of uncertainty are added in quadrature.

\begin{figure}[ht]
\begin{center}
\begin{tabular}{cc}
  \epsfig{figure=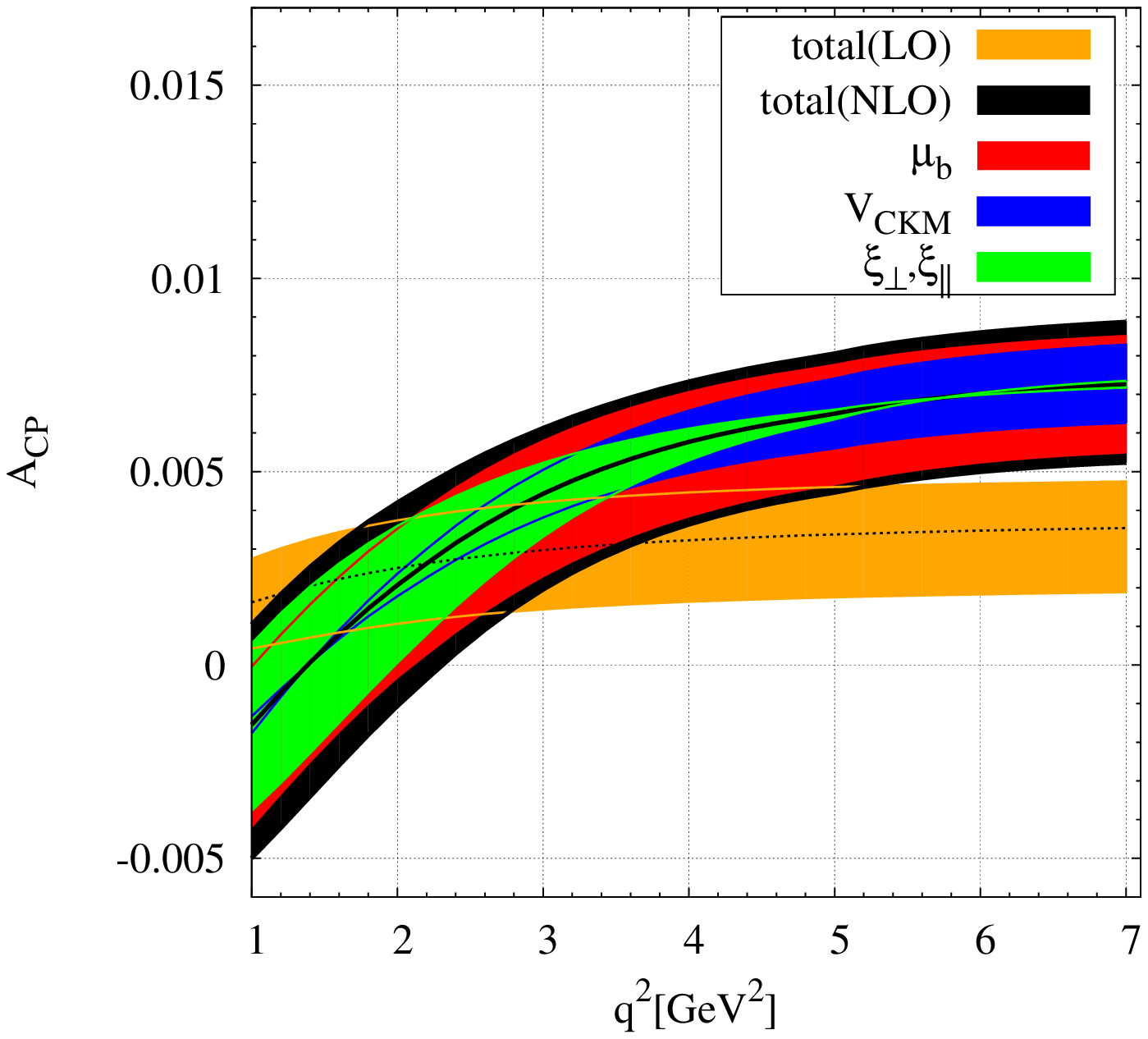, height=6.5cm, angle=0} &
  \epsfig{figure=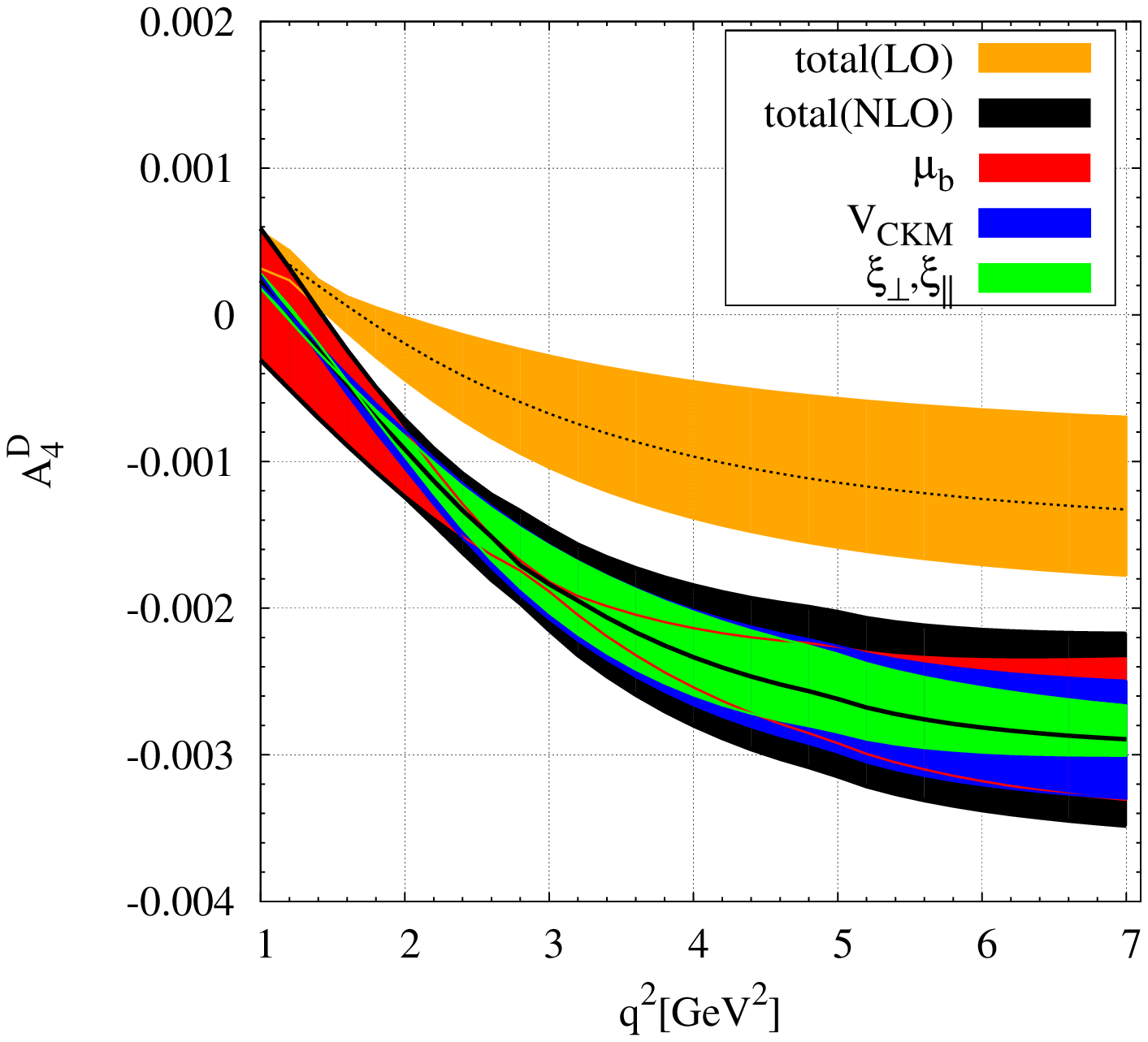, height=6.5cm, angle=0} \\
[-15mm]
  \epsfig{figure=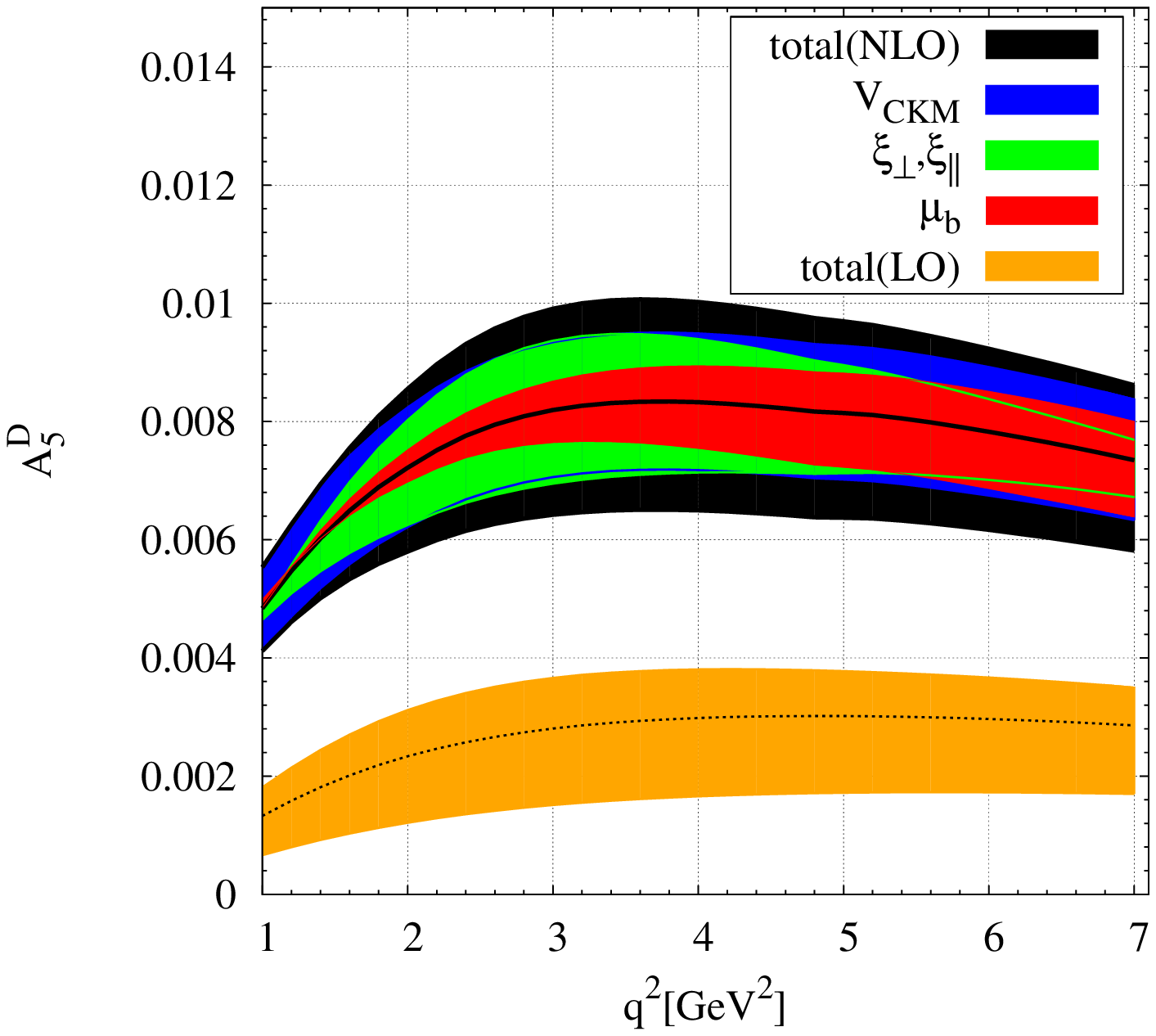, height=6.5cm, angle=0} &
  \epsfig{figure=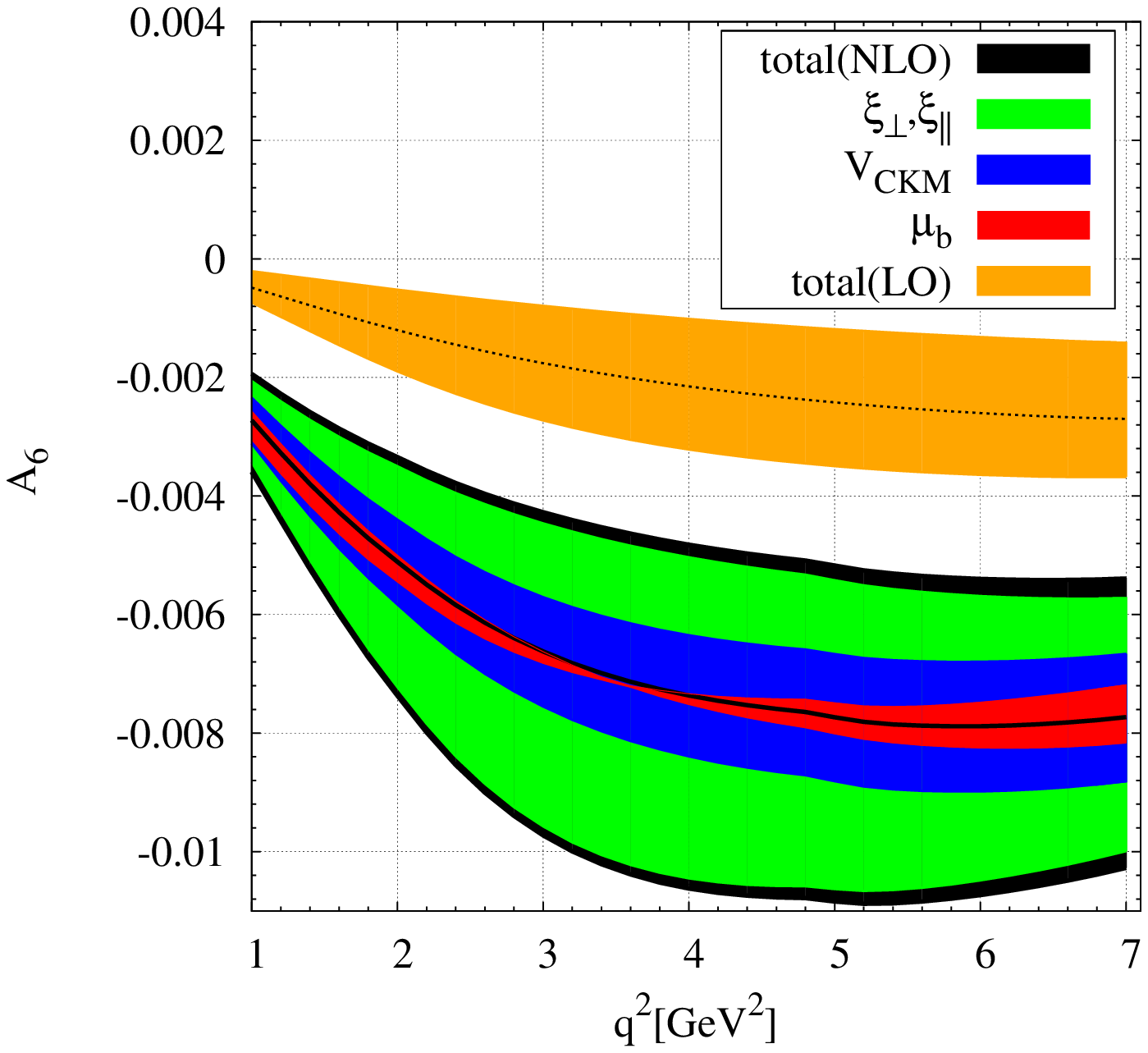, height=6.5cm, angle=0} \\
[-15mm]
  \epsfig{figure=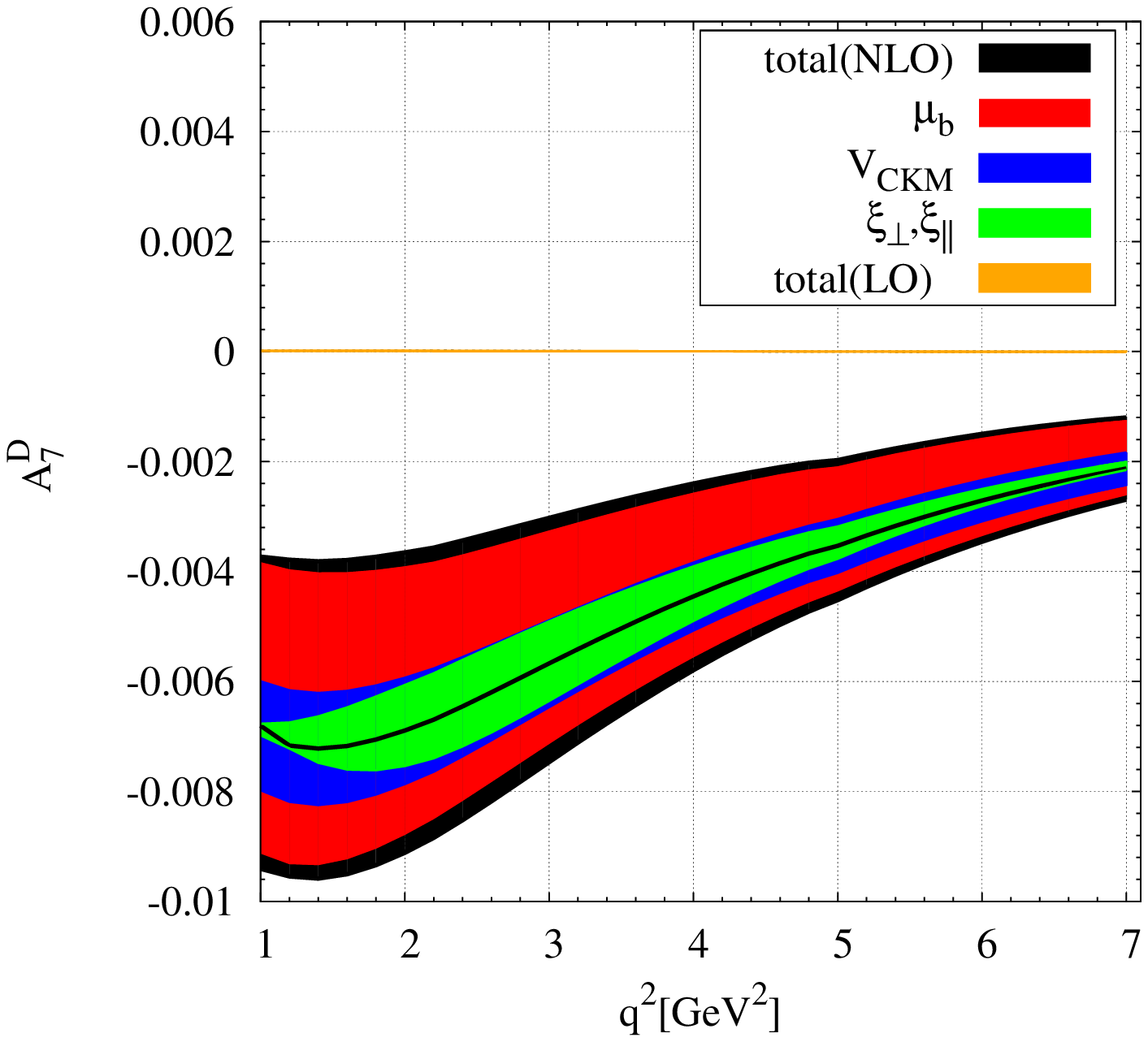, height=6.5cm, angle=0} &
  \epsfig{figure=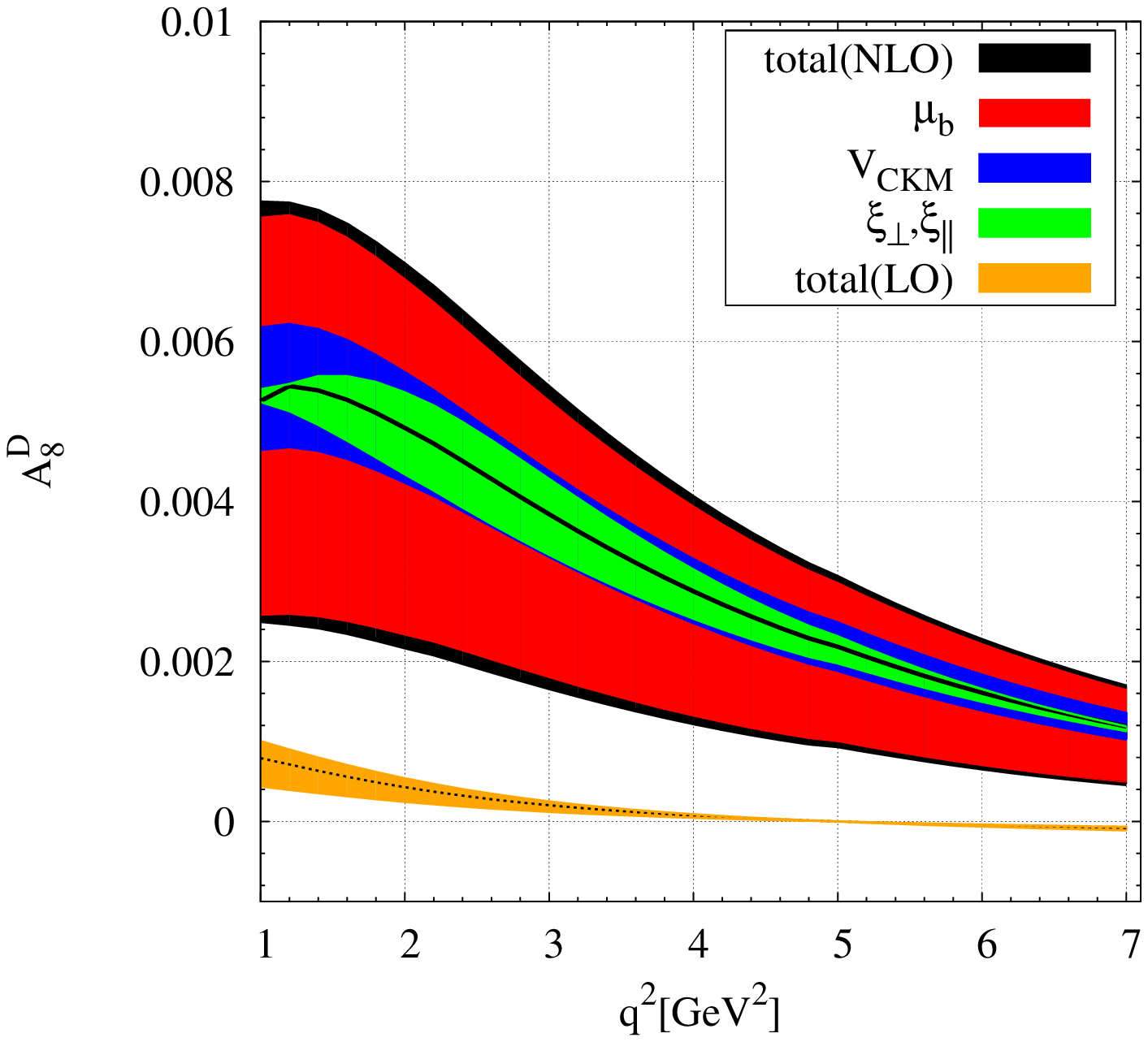, height=6.5cm, angle=0} 
\\ [-14mm]
\end{tabular}
\end{center}
\caption{ \label{fig:A78sm:q2} 
  The CP asymmetries $A_{\rm CP, 6}(q^2)$ and $A_{4,5,7,8}^{D}(q^2)$ in the SM 
  in the low-$q^2$ region at LO and NLO in QCDF.
  The various bands show the uncertainty due to the form 
  factors, the CKM parameters and $\mu_b$ separately, whereas the overall band 
  indicates the total uncertainty. }
\end{figure}

In \reffig{fig:A78sm:q2} we show the T-odd CP asymmetries $A_{7,8}^D$  and 
the T-even ones $A_{\rm CP}, A_{4,5}^D$ and $A_6$ as a function of $q^2$. 
The various bands indicate the uncertainties due to the form factors, 
the CKM parameters, $\mu_b$ and the total uncertainty. The asymmetries 
$A_3$ and $A_9$ are not shown, since they vanish in the SM at lowest 
order in $1/E$. (A small finite value is induced by the 
strange quark mass.) Hence, their leading contributions may arise as
\begin{equation}
  A_{3,9} \sim 
  \im [\hat{\lambda}_u] \, {\cal{O}}(\LamConf/E) \sim {\cal{O}}(10^{-3}).
\end{equation}

The LO predictions for the CP asymmetries are also included in 
\reffig{fig:A78sm:q2}. The higher order  $\alS$-corrections 
increase the size of the CP asymmetries. For $A_7^D$ and $A_8^D$
this happens because their respective LO values are suppressed 
by cancellations. Specifically, in the SM 
\begin{align}
  \label{eq:A7:cancel}
  A_7^{D} & \sim \im [\hat{\lambda}_u] \re \left[ 
    \frac{{\cal T}_\perp^{(u)}}{\xi_\perp}\ + 
     \frac{q^2}{M_B^2} \frac{{\cal T}_\parallel^{(u)}}{\xi_\parallel} \right], & 
\end{align}
which vanishes at LO in QCDF, see \refeq{eq:calT:LO}, and also \cite{FK2003}. 
(Our value of $A_7^D$ at LO is tiny but finite since in the numerical 
analysis we do not neglect kinematical factors $M_{K^*}^2/M_B^2$.)
The asymmetry $A_8^D$ is subject to similar cancellations, although here
an additional LO term exists, which is, however, numerically subleading.
The values of $A_7^D$ and  $A_8^D$ are therefore determined
by the NLO $\alS$-corrections resulting in a large $\mu_b$ uncertainty.
The impact of the higher order terms
on the T-even asymmetries is sizeable, but less pronounced.
We discuss further details of the SM CP asymmetries
in the context of the integrated CP asymmetries.

%%%%%%%%%%%%%%%%%
%\TABLE[bt]{
\begin{table}
\centering
\begin{tabular}{||c|c c c|c||c||}
\hline \hline
  & SM $\cdot 10^{-3}$ & $\xi_{\perp, \parallel} [\%]$ &  $\mu_b [\%]$
& SM LO $\cdot 10^{-3}$ & SM$(B^\mp)$ $\cdot 10^{-3}$
\\
\hline
\hline
\multirow{2}{*}{$\langle A_{\rm CP} \rangle$}
& ${4.2}_{ - 2.5}^{ + 1.7}$ & ${}_{ - 24}^{ + 19}$ & ${}_{ - 51}^{ + 33}$
& ${3.0}_{ - 1.5}^{ + 1.2}$ 
& ${10.0}_{ - 2.9}^{ + 2.3}$
\\[0.5ex]
& ${4.8}_{ - 2.4}^{ + 1.7}$ & ${}_{ - 17}^{ + 13}$ & ${}_{ - 44}^{ + 29}$
& ${3.1}_{ - 1.6}^{ + 1.2}$
& ${9.9}_{ - 2.8}^{ + 2.2}$
\\[0.5ex]
\hline
\multirow{2}{*}{$\langle A_4^D \rangle$}
& ${-1.8}_{ - 0.3}^{ + 0.3}$ & ${}_{ - 8}^{ + 11}$ & ${}_{ - 6}^{ + 2}$
& ${-0.7}_{ - 0.4}^{ + 0.4}$
& ${-0.7}_{ - 0.3}^{ + 0.4}$
\\[0.5ex]
& ${-2.0}_{ - 0.4}^{ + 0.4}$ & ${}_{ - 8}^{ + 11}$ & ${}_{ - 8}^{ + 7}$
& ${-0.8}_{ - 0.4}^{ + 0.5}$
& ${-1.1}_{ - 0.4}^{ + 0.4}$
\\[0.5ex]
\hline
\multirow{2}{*}{$\langle A_5^D \rangle$}
& ${7.6}_{ - 1.6}^{ + 1.5}$ & ${}_{ - 13}^{ + 10}$ & ${}_{ - 8}^{ + 7}$
& ${2.7}_{ - 1.2}^{ + 0.8}$
& ${10.0}_{ - 2.3}^{ + 2.2}$
\\[0.5ex]
& ${7.6}_{ - 1.6}^{ + 1.5}$ & ${}_{ - 12}^{ + 9}$ & ${}_{ - 9}^{ + 7}$
& ${2.7}_{ - 1.2}^{ + 0.8}$
& ${9.8}_{ - 2.1}^{ + 2.1}$
\\[0.5ex]
\hline
\multirow{2}{*}{$\langle A_6 \rangle$}
& ${-6.4}_{ - 2.7}^{ + 2.2}$ & ${}_{ - 39}^{ + 31}$ & ${}_{ - 2}^{ + 0}$
& ${-1.9}_{ - 0.9}^{ + 1.0}$
& ${-6.3}_{ - 2.6}^{ + 2.1}$
\\[0.5ex]
& ${-6.7}_{ - 2.7}^{ + 2.2}$ & ${}_{ - 37}^{ + 30}$ & ${}_{ - 3}^{ + 1}$
& ${-2.0}_{ - 1.0}^{ + 1.1}$
& ${-6.6}_{ - 2.7}^{ + 2.2}$
\\[0.5ex]
\hline
\multirow{2}{*}{$\langle A_7^D \rangle$}
& ${-5.1}_{ - 1.6}^{ + 2.4}$ & ${}_{ - 8}^{ + 11}$ & ${}_{ - 26}^{ + 42}$
& \multirow{2}{*}{$< 10^{-2}$ } 
& ${-7.1}_{ - 1.9}^{ + 2.6}$
%${0.005}_{ - 0.002}^{ + 0.002}$
\\[0.5ex]
& ${-4.6}_{ - 1.4}^{ + 2.1}$ & ${}_{ - 6}^{ + 10}$ & ${}_{ - 25}^{ + 42}$
&
& ${-6.5}_{ - 1.7}^{ + 2.3}$ 
%${0.003}_{ - 0.002}^{ + 0.001}$
\\[0.5ex]
\hline
\multirow{2}{*}{$\langle A_8^D \rangle$}
& ${3.5}_{ - 2.0}^{ + 1.4}$ & ${}_{ - 10}^{ + 7.4}$ & ${}_{ - 53}^{ + 37}$
& ${0.2}_{ - 0.08}^{ + 0.04}$
& ${3.4}_{ - 2.0}^{ + 1.4}$
\\[0.5ex]
& ${3.1}_{ - 1.7}^{ + 1.3}$ & ${}_{ - 10}^{ + 6}$ & ${}_{ - 53}^{ + 37}$
& ${0.14}_{ - 0.06}^{ + 0.03}$
& ${3.1}_{ - 1.8}^{ + 1.3}$
\\[0.5ex]
\hline
\multirow{2}{*}{$\langle A_{3,9} \rangle^\dagger$}
&
\multirow{2}{*}{$\order{1}$} & & &
\multirow{2}{*}{$\order{1}$ }& \multirow{2}{*}{$\order{1}$ }\\[0.5ex]
 & & & &&
\\[0.5ex]
\hline \hline
\end{tabular}
\caption{ \label{tab:CPas:SM:res} SM predictions for the integrated CP
  asymmetries in units of $10^{-3}$ with the integration boundaries
  $(q^2_{\rm min}, q^2_{\rm max}) = (1, 6), (1, 7) \GeV^2$ 
  (from top to bottom).
  We take into account uncertainties from the form factors,
  the scale dependence  $\mu_b$ and the CKM parameters, all of them
  added in quadrature for the total uncertainty. 
%  The form factor uncertainty employed is
%  $11\%$ and $14\%$ for $\xi_\perp$ and $\xi_\parallel$,
%  respectively, and $\mu_b$ is varied within $[m_b/2, 2 m_b]$.
  The relative uncertainties due to  $\xi_{\perp},\xi_\parallel$ and  $\mu_b$
  are also shown separately. The asymmetries at LO in $\alS$ and the NLO ones 
  for charged $B$-decays are given as well, see text for details.
  ${}^\dagger$The leading contributions $\langle A_{3,9} \rangle$
  in the SM are power counting estimates only. 
}
\end{table}
%%%%%%%%%%%%%%%%%

We find that all $q^2$-integrated CP asymmetries  $\langle A_i^{(D)} \rangle$ 
are less than $ \order{10^{-2}}$ in the SM. This 
can be seen in \reftab{tab:CPas:SM:res}, where we give the results for the 
two cuts $(q^2_{\rm min}, q^2_{\rm max}) = (1, 6) \GeV^2$ (upper entries) 
and $(1, 7) \GeV^2$ (lower entries), respectively. The uncertainties from 
the form factors and $\mu_b$ are also shown separately.
Due to the overall CKM factor $\im [ \hat{\lambda}_u ] $, 
all CP asymmetries suffer
from a universal $15 \%$ uncertainty related to that.
The corresponding values for the NLO CP averaged decay rates are
$\langle N_\Gamma \rangle /2 = (1.1 \pm 0.3) \cdot 10^{-19} \, \mbox{GeV}$ and 
$\langle N_\Gamma \rangle /2 =(1.3^{+0.4}_{-0.3}) \cdot 10^{-19}\, \mbox{GeV}$ 
for $(q^2_{\rm min}, q^2_{\rm max}) = (1, 6) \GeV^2$ and $(1, 7) \GeV^2$, 
respectively, assuming ${\cal{B}}(K^* \to K \pi)=100 \%$. 
The uncertainty in the rate is about 25 \%  from the form factors, 
7 \% from $V_{ts}$ and order one percent from $\mu_b$. 

The form factor induced uncertainty in the asymmetries depends on the amount 
of cancellations between the numerator and the decay rate in 
the denominator. We recall that we vary the two form factors within their 
uncertainties independently. 
Taking into account correlations would reduce the errors in the ratios, but 
requires control over the parameters  
of the LCSR calculation \cite{Ball:2004rg}, which is beyond the scope of 
this work.  Since the decay rate 
is dominated by the longitudinal $K^*$ polarization driven by $\xi_\parallel$, 
see the discussion following \refeq{eq:dGdq:ff}, the strongest form factor 
uncertainty is seen in $A_6$ being proportional to $\xi_\perp^2$. 
The other asymmetries $A_i^D$, $i=4,5,7,8$ with numerator 
$\propto \xi_\parallel \xi_\perp$ receive more efficient cancellations.

As can be seen from \reftab{tab:CPas:SM:res}, $\langle A_{\rm CP} \rangle$, 
$\langle A_7^D \rangle$ and 
$\langle A_8^D \rangle $ exhibit a massive $\mu_b$ dependence of 
order 50 \%. The CP asymmetries $A_i^{(D)}$ with $i = 4,5,6$ are not subject
to the cancellations mentioned after \refeq{eq:A7:cancel} and have a smaller 
residual $\mu_b$ uncertainty below ten percent. The $\mu_b$ dependence of 
$\langle A_6 \rangle$ of a few percent is accidentally small due to 
significant cancellations between different $q^2$-regions, see the crossing
of the $\mu_b$ bands in $A_6$ near $q^2 \simeq (3-4) \, \mbox{GeV}^2$ 
in \reffig{fig:A78sm:q2}. 

Furthermore, we study the impact of higher order contributions in QCDF on the 
$\langle A_i^{(D)} \rangle$. The shift from LO in $\alpha_s$ to NLO
is substantial. Switching off the spectator interactions reduces the size of
$\langle A_{7,8}^D \rangle$ by about 10 \%, and by less for the other 
asymmetries. We also give in \reftab{tab:CPas:SM:res} the NLO SM 
predictions for charged $B$-decays. The splitting between
the CP asymmetries in neutral versus charged $B$-decays is dominated by
weak annihilation contributions from current-current operators and
varies a lot in size:
$\langle A_{5,7}^{D} \rangle$ ($\langle A_{ \rm CP} \rangle$)
increase by ${\cal{O}}(30 \%)$ (a factor of two) from neutral to charged 
$B$-decays, whereas $\langle A_4^{D} \rangle$ decreases by $\sim 1/2$. The
splitting for $\langle A_{6,8}^{(D)} \rangle$ is at the few percent level.

The SM predictions for the untagged, time-integrated CP asymmetries
$\langle A_i^{(D) mix} \rangle$, $i=5,6,8,9$, in 
$\bar B_s, B_s \to \phi (\to K^+ K^-) \bar l l$ decays  
\refeq{eq:Aimix} can be inferred from the $\langle A_i^{(D)} \rangle$ 
in neutral $B_d$-decays, which are given in \reftab{tab:CPas:SM:res}.
Corrections arise from $SU(3)$ flavor breaking, which is expected to be small
in the ratios, from the $B_s$-width difference at the level of ten percent 
and from spectator interactions. All these effects are subdominant with 
respect to the theoretical uncertainties of the SM predictions.

%--------+---------+---------+---------+---------+---------+---------+---------+
\section{Beyond the Standard Model \label{sec:NP}}

This section contains the model-independent analysis of the CP asymmetries. 
We consider NP contributions to
the operators $\Op_{7,9,10}$ which are part of the effective Hamiltonian 
\refeq{eq:Heff} of the SM, as well as NP contributions to the chirality 
flipped ones $\Op'_{7,9,10}$. We allow the respective NP coefficients
$C_i^{\rm NP}$ and $C_i^{ ' \rm NP}=C_i'$ for 
$i = 7,9,10$ to vary in magnitude and phase, denoted by $\phi_i$, 
within the constraints from the FCNC $B$-decay data.
The radiative and semileptonic $b \to s$ transitions are the most 
important ones for our analysis. The relevant data and SM predictions 
are given in \reftab{tab:exp:data}. 

\begin{table}
%\TABLE[ht]{
\renewcommand{\arraystretch}{1.1}
\centering
\begin{tabular}{||l|c|c||}
\hline \hline
  observable & SM & data\\
\hline \hline
  $\BR(\BXsgamma)^a$ &
  $(3.15 \pm 0.23) \cdot 10^{-4}$ \cite{Misiak:2006zs}&
  $(3.52 \pm 0.25) \cdot 10^{-4}$ 
  \hfill \cite{Barberio:2006bi} 
\\ 
  $S_{K^* \gamma}^b$ & 
  $(-2.8^{+0.4}_{-0.5}) \cdot 10^{-2}$ & 
  $-0.19 \pm 0.23$ \hfill \cite{Barberio:2006bi,SKstargamma} 
\\
\hline
  $\BRll{\BXsll}{1,6}{\mbox{}}$ &
  $(1.59 \pm 0.11) \cdot 10^{-6}$ \cite{Huber:2005ig}  &
  $ (1.60 \pm 0.51) \cdot 10^{-6}$ 
  \hfill \cite{Aubert:2004it}
\\
  $\BRll{\BXsll}{>0.04}{\mbox{}}$ &
  $(4.15 \pm 0.70) \cdot 10^{-6}$ \cite{Ali:2002jg}  &  
  $ (4.5 \pm 1.0) \cdot 10^{-6}$ 
  \hfill \cite{Yao:2006px}
\\
$\langle A_{\rm FB} \rangle_{[high \, q^2]}^c $ & $<0$ & 
$-(0.76^{+0.52}_{-0.32} \pm 0.07)$  \hfill 
\cite{Aubert:2006vb}, \hfill also \cite{Abe:2004ir,Ishikawa:2006fh} \\
\hline 
  $\BR(\Bstomm)$ & 
  $\simeq 3 \cdot 10^{-9}$ &
  $ < 4.7 \cdot 10^{-8}$ 
  \hfill at $90 \%$ C.L. \cite{:2007kv} \\
\hline \hline

\end{tabular}
\caption{\label{tab:exp:data} Relevant $b \to s \gamma$ and
$b \to s \bar l l$ observables. 
${}^a$With photon energy cut $E_\gamma> 1.6$ GeV.
${}^b$SM value obtained with $m_s=0.12 \GeV$.
${}^c$Note the different lepton angle convention between 
\cite{Ishikawa:2006fh,Aubert:2006vb} and this work.}
\end{table}

In our analysis the NP Wilson coefficients are leading order coefficients. 
All Wilson coefficients are understood as evaluated at the low, $\mu_b$-scale.
We start with a discussion of the experimental constraints.

\subsection{Experimental Constraints}

The radiative decays induced by $b \to s \gamma$ probe the 
electromagnetic dipole coefficients $C_7^{}$ and $C'_7$. In the NP scenarios 
considered here, the flipped dipole coefficient has no interference terms 
in the radiative decay rates. Hence, these observables constrain 
only the magnitude of $C_7'$ and not its phase. 

We take into account the $\BXsgamma$ branching ratio for which we adopt the 
NNLO SM results from \cite{Misiak:2006zs}. To account for the
missing higher order calculation of the beyond-the-SM amplitude, 
we take for the theoretical uncertainty of 
the NP contribution twice the SM uncertainty. We apply the experimental 
constraints at 90 $\%$ C.L.
We checked that the direct CP asymmetry in $\BXsgamma$, e.g., 
\cite{Hurth:2003dk}, does not give constraints beyond those from the 
$\BXsgamma$ branching ratio. 

The time-dependent CP asymmetry 
$S_{K^* \gamma}$ in $\bar B_d, B_d \to K^{*0} ( \to K^0 \pi^0) \gamma$ 
\cite{Atwood:1997zr} is important since it is sensitive to the interference 
of photons with different polarization, that is, photons coming from 
${\cal{O}}_7$ versus ${\cal{O}}_7'$. To illustrate the dependence on
the Wilson coefficients, we give $S_{K^* \gamma}$ at lowest order 
(indicated by the superscript $(0)$ for the contributions already 
present in the SM): 
\begin{equation} \label{eq:SKstg}
S_{K^* \gamma} =-\frac{ 2 |r| }{1+ |r|^2} 
\sin \left(2 \beta -\arg (C_7^{(0)} C_7') \right) , ~~~~
r=C_7'/C_7^{(0)} \, .
\end{equation}
Here we assume that there is no beyond-the-SM physics in 
$B_d -\bar B_d$-mixing, and its phase is given by the CKM matrix elements.
We calculate the exclusive $\bar B \to \bar K^* \gamma$ decay amplitude with 
QCD factorization following \cite{Beneke:2004dp} including
$\alpha_s$-corrections. The constraints from 
$S_{K^* \gamma}$ exclude some regions with $|r|$ of order one, 
unless the CP phases conspire
to suppress the sine  in \refeq{eq:SKstg}, see below.

The second class of constraints stems from the semileptonic 
transitions and applies to all Wilson coefficients we consider, 
$C_{7,9,10}^{(')}$.
The inclusive $\BXsll$ decays can be predicted with high accuracy, 
in the low-$q^2$ region at the level of $\lesssim 10\%$ \cite{Huber:2005ig}, 
but also the high-$q^2$ region is theoretically accessible.
As can be seen in \reftab{tab:exp:data}, we utilize the integrated branching 
ratios in the low-$q^2$ region with $q^2 \in [1, 6] \GeV^2$, 
$\BRll{\BXsll}{1,6}{\mbox{}}$, as
well as for $q^2 > 0.04 \GeV^2$, $\BRll{\BXsll}{>0.04}{\mbox{}}$.
The latter has been experimentally obtained by cutting out events
with $q^2$ close to the first and second charmonium resonance, hence bears 
some model-dependence. We use the corresponding theory predictions from
\cite{Huber:2005ig} and  \cite{Ali:2002jg}, respectively.
The decay distributions with NP are given in \cite{Guetta:1997fw}.
The treatment of uncertainties is as for the $\BXsgamma$ branching 
ratio.

Concerning the exclusive $\BtoKll$ and $\BtoKastll$ decays, 
we do not use the branching ratios for our model-independent analysis:
the constraints are in general weaker than the ones from ${\cal{B}}(\BXsll)$ 
due to the larger theoretical and experimental uncertainties.
A particular difficulty with the available exclusive semileptonic 
decay data is the presence of measurements with different
dilepton mass cuts, some of which in addition include regions 
where QCDF or SCET does not apply.

We employ instead early data on the $\BtoKastll$ forward-backward asymmetry 
from Belle and BaBar \cite{Abe:2004ir,Ishikawa:2006fh,Aubert:2006vb}. 
While these measurements have large uncertainties, both experiments 
strongly support the sign of  $A_{\rm FB}$ in the high-$q^2$ region above the 
second charmonium peak to be SM-like.

A rigorous theory calculation of the exclusive $\BtoKastll$ decays in this 
kinematical region can be facilitated with an operator product expansion 
in $\LamConf/Q$ and $m_c^2/Q^2$ where $Q = \{\sqrt{q^2},  m_b\}$ put 
forward in \cite{Grinstein:2004vb}. The leading contribution and also the 
order $m_c^2/Q^2$ terms do not introduce new non-perturbative matrix elements 
beyond naive factorization. Corrections start to enter 
at $\order{\alS \LamConf/Q}$. The framework holds at low recoil, 
$(M_B - M_{K^*})^2 - 2 M_B \LamConf \lesssim q^2 < (M_B - M_{K^*})^2$, which 
covers the large dilepton mass region above the $\Psi'$ resonance,
$q^2 \gtrsim 14 \GeV^2$.

To leading order in the $1/Q$-expansion we obtain $A_{\rm FB}$ at low recoil as
\begin{align} \label{eq:AFB-OPE}
A_{\rm FB}(q^2) \propto  \re\Big[
(C_9^{\rm eff}(q^2)+\frac{2m^2_b}{q^2}C_7^{\rm eff})\,C^*_{10}
-(C'_9+\frac{2m^2_b}{q^2}C'_7)\,C^{'*}_{10}\Big] .
\end{align}
The effective coefficients read as
$C_9^{\rm eff}(q^2)=C_9^{}+(4/3 \, C_1^{}+C_2^{}) g(q^2) + \ldots$ 
and $C_7^{\rm eff }=C_7^{}+\ldots$, where
$4/3 \, C_1^{}+C_2^{} \simeq 0.61$ are the dominant SM coefficients.
The full expressions including the higher order $\alpha_s$-corrections and 
the QCD penguin contributions are given in \cite{Grinstein:2004vb}
and are included in our numerical analysis.
The lowest order charm loop function is given as 
\begin{eqnarray}
g(q^2)= \frac{8}{27} +\frac{4}{9} \left( 
\ln \frac{\mu^2}{q^2} + i\pi \right) ,
\end{eqnarray}
which agrees with the perturbative quark loop function for massless quarks.
Interestingly, the dependence on form factors can be factored out in 
$A_{\rm FB}$ \refeq{eq:AFB-OPE} at this order. 
We require then the sign of $\langle A_{\rm FB} \rangle $ integrated over
$q^2 > 14 \GeV^2$ to be negative.

 \begin{figure}
 \begin{center}
\epsfig{figure=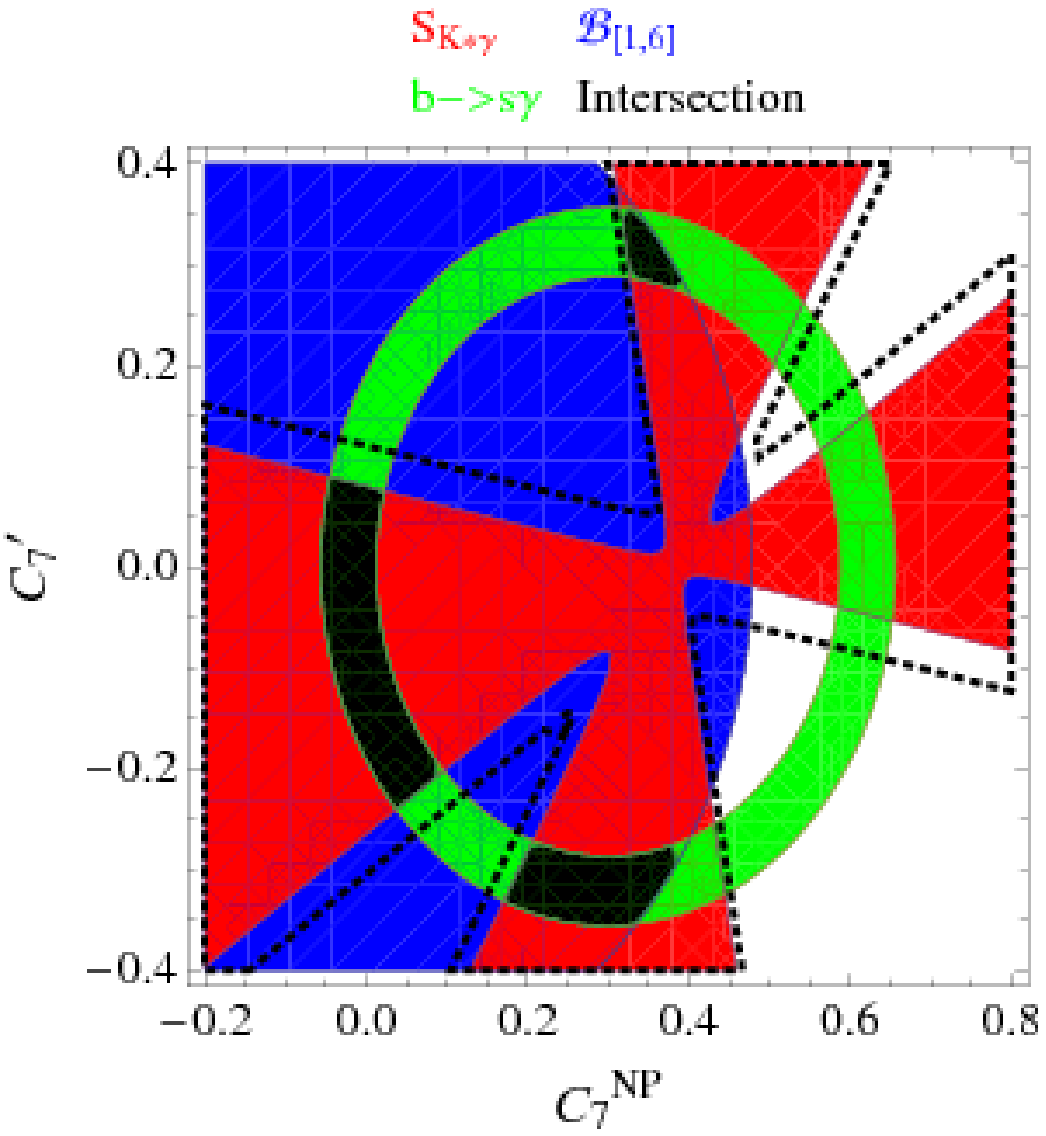, height=6cm, angle=0} 
 \epsfig{figure=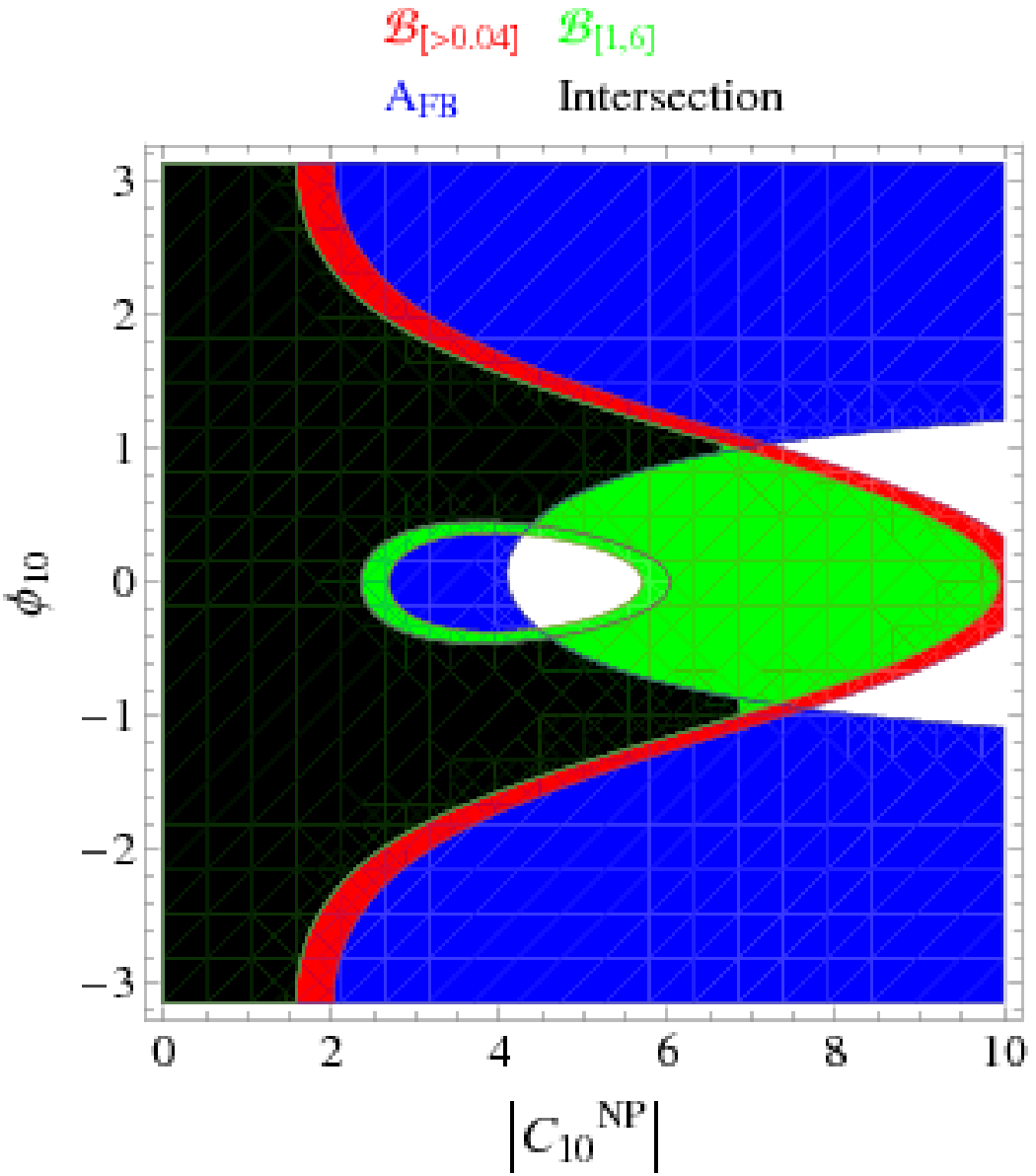, height=6cm, angle=0}
 \end{center} \caption{ 
\label{fig:wilsonNP}
Allowed regions of Wilson coefficients in specific NP scenarios 
after applying the experimental constraints from radiative and 
semileptonic $b \to s$ processes
as indicated. The black areas are allowed by all constraints.
In the left panel we show $C_7'$ versus
$C_7^{\rm NP}$ assuming real Wilson coefficients. 
We give the magnitude of $C_{10}^{\rm NP}$ versus its CP phase $\phi_{10}$
in the right panel.
In both plots all other NP Wilson coefficients have been set to zero.
For details see text.}
 \end{figure}

We show the impact of the FCNC constraints on the NP Wilson coefficients 
for two NP scenarios in \reffig{fig:wilsonNP}.
The areas allowed by all constraints are given in black.
We learn that the observables (each shown in a different color) 
yield complementary information, and that the SM
is allowed, as well as many significantly different NP solutions.

In the left plot, we entertain NP only in $C_7$ and $C_7'$, and
assume further no NP CP phases. The regions allowed by
${\cal{B}}(\BXsgamma)$, $S_{K^* \gamma}$  and
 $\BRll{\BXsll}{1,6}{\mbox{}}$ are shown as the green ring, the red cross and
the blue half circle, respectively. The impact of 
$S_{K^* \gamma}$ is significant. 
The semileptonic decay excludes in this NP scenario the flipped sign solution
for the photonic dipole coefficient $C_7 \simeq -C_7^{\rm SM} \sim 0.31$. 
Note that dimensional analysis suggests that 
power corrections to $r$ of the order $C_2 \LamConf/(3 m_b C_7) \sim 0.1$ 
may induce a larger SM contribution to $S_{K^* \gamma}$ 
than ${\cal{O}}(m_s/m_b)$ \cite{Grinstein:2004uu}.
We show the resulting region in the $C_7'-C_7^{\rm NP}$-plane 
by the dashed lines in the left plot of \reffig{fig:wilsonNP}.
Because of the present experimental situation, however,
the inclusion of the power corrections corresponds only to a small 
enlargement of the allowed parameter space.
Note also that Ref.~\cite{Ball:2006eu} estimated the non-perturbative 
SM contributions 
to be smaller than the ones coming from naive power counting.

In the right plot we allow only for NP in $C_{10}$, and show the allowed 
regions in $|C_{10}^{\rm NP}|$ and the CP phase $\phi_{10}$.
Fixing the sign of $A_{\rm FB}$ (blue) and the semileptonic branching ratios 
$\BRll{\BXsll}{1,6}{\mbox{}}$ (green) and
$\BRll{\BXsll}{>0.04}{\mbox{}}$ (red) yield orthogonal constraints. 
An upper bound on the magnitude of $C_{10}^{\rm NP}$ is obtained
with the aid of $A_{\rm FB}$ as $|C_{10}^{\rm NP}| \lesssim 7$,
improving on the bound from the branching ratios alone,
$|C_{10}^{\rm NP}| \lesssim 10$.

%--------+---------+---------+---------+---------+---------+---------+---------+
\subsection{CP Asymmetries with New Physics \label{sec:NP:CPasy}}

%\TABLE[bt]{
\begin{table}
\centering
\begin{tabular}{||c|cccc||}
\hline \hline
& generic NP
& $C_{10}^{\rm NP}$ only & $C_{10}^{' \rm NP}$ only &$C_9^{\rm NP}$ only
\\
\hline
\hline
$\langle A_{\rm CP} \rangle$
& $[-0.12, 0.10]$
& $[3, 8]\cdot 10^{-3}$
& SM-like
%$[2, 6]\cdot 10^{-3}$
&$[-0.02, 0.02]$
\\[0.5ex]
$\langle A_3 \rangle$
& $[-0.08, 0.08]$
& SM-like
& SM-like
& SM-like
\\[0.5ex]
$\langle A_4^D \rangle$
& $[-0.04, 0.04]$
& $[-4, -1] \cdot 10^{-3} $
&$[-3, -1]\cdot 10^{-3}$
&$[-0.01, 0.01]$
\\[0.5ex]
$\langle A_5^D \rangle$
& $[-0.07, 0.07]$
& $[-0.04, 0.04]$
& $[-0.02, 0.04]$
&$[5, 9]\cdot 10^{-3}$
\\[0.5ex]
$\langle A_6 \rangle$
& $[-0.13, 0.11]$
& $[-0.05, 0.05]$
&$[-9, -3]\cdot 10^{-3}$
& SM-like
%$[-8, -4]\cdot 10^{-3}$
\\[0.5ex]
$\langle A_7^D \rangle$
& $[-0.76, 0.76]$
& $[-0.48, 0.48]$
& $[-0.38, 0.38]$
& SM-like
%$[-6, -3]\cdot 10^{-3}$
\\[0.5ex]
$\langle A_8^D \rangle$
& $[-0.48, 0.48]$
& $[2, 7] \cdot 10^{-3}$
& $[-0.28, 0.28] $
& $[-0.17, 0.17] $
\\[0.5ex]
$\langle A_9 \rangle$
& $[-0.62, 0.60]$
& SM-like
& $[-0.20, 0.20]$
& SM-like
\\[0.5ex]
\hline
$\BR(\Bstomm)$
& $<1.4 \cdot 10^{-8}$
& $<6.3 \cdot 10^{-9}$
& $<1.3 \cdot 10^{-8}$
& SM
\\[0.5ex]
\hline \hline
\end{tabular}
\caption{ \label{tab:CPas:NP:res}
The ranges of the integrated CP asymmetries $\langle A_i^{(D)} \rangle$
for
$(q^2_{\rm min}, q^2_{\rm max}) = (1, 6) \GeV^2$ after applying the
experimental constraints at 90\% C.L.~for the generic NP scenario and
those with NP in $C_{10}, C_{10}'$ or $C_9$ only. The upper limits on
$\BR(\Bstomm)$ are also shown. For details see text. }
\end{table}
%}

The dependence of the CP asymmetries $A_{i}^{(D)}$ on the Wilson 
coefficients can be seen from the
analytical (NLO) formulae in \refapp{app:CP:asy}. 
We also provide numerical 
model-independent formulae for the $\BtoKKpill$ branching ratio and 
CP asymmetries in \refapp{app:num:CP:asy}. The
numerators of $A_{{\rm CP},3,4}^{(D)}$ are sensitive to
$C_{7,9}$ and $C'_{7,9}$ whereas the numerators of $A_{5,7}^{D}$ 
and $A_6$ probe $C_{7,10}$ and $C'_{7,10}$.  
The numerators of $A_{8,9}^{(D)}$ can be affected by all Wilson 
coefficients considered here. Recall also that $A_{3,9}$ are 
very sensitive to the 
flipped Wilson coefficients since $A_{3,9}$ 
vanish in the limit  $C'_i \to 0$ at lowest order in the $1/E$-expansion.

To see directly these features of the T-odd asymmetries, we provide LO 
formulae:
\begin{align}
  A_7^D & =
 2 {\cal A}^D   \frac{ \hat m_b}{\hat s} ( 1- \hat s)
\im \left[(C_{10}^{(0)} - C_{10}') (C_7^{\rm eff (0)} - C_7')^* \right], 
\\[2ex]
  A_8^D & = {\cal A}^D \beta_l \bigg\{
    \im \left[ C_9^{(0)} C_9'^{*} + C_{10}^{(0)} C_{10}'^{*} 
             + \frac{4 \hat{m}_b^2}{\hat{s}} C_7^{\rm eff (0)} C_7'^{*} \right.
\nonumber\\             
    & \hspace{2.5cm} \left.
      + \frac{\hat{m}_b}{\hat{s}} \big( (1 - \hat{s}) (C'_7 C_9'^{*} - C_7^{\rm eff (0)} C_9^{(0)*} )
                  + (1 + \hat{s}) (C_7^{\rm eff (0)} C_9'^{*} - C'_7 C_9^{(0)*})\big)
        \right]
\nonumber\\
   & \hspace{1.3cm} - \re (Y^{(0)}) \im 
\left[ C'_9 + \frac{\hat{m}_b}{\hat{s}} 
\big( (1 - \hat{s}) C_7^{\rm eff (0)}  + (1 + \hat{s}) C'_7 \big) \right]
    \bigg\}  + {\cal{O}}(\hat \lambda_u), 
 \\[2ex]
   A_9 & =  4 {\cal A}^D \beta_l \bigg\{
     \im \left[ C_9^{(0)} C_9'^{*} + C_{10}^{(0)} C_{10}'^{*} 
              + \frac{4 \hat{m}_b^2}{\hat{s}^2} C_7^{\rm eff (0)} C_7'^{*} 
              + \frac{2 \hat{m}_b}{\hat{s}} (C_7^{\rm eff (0)} C_9'^{*} - 
C'_7 C_9^{(0)*})
         \right]
 \nonumber\\
    & \hspace{0.5cm}
      - \frac{\hat{m}_b}{\hat{s}} \re (Y^{(0)}) \im \left[ 
        2 C'_7 + \frac{\hat{m}_b}{\hat{s}} C'_9 \right]
    \bigg\} + {\cal{O}}(\hat \lambda_u) ,
\end{align}
where for $A_8^D ,A_9$ we neglected SM CP violation suppressed by 
$\hat \lambda_u$, see \refapp{app:CP:asy} for details.

 \begin{figure}
 \begin{center}
  \begin{tabular}{cc}
\epsfig{figure=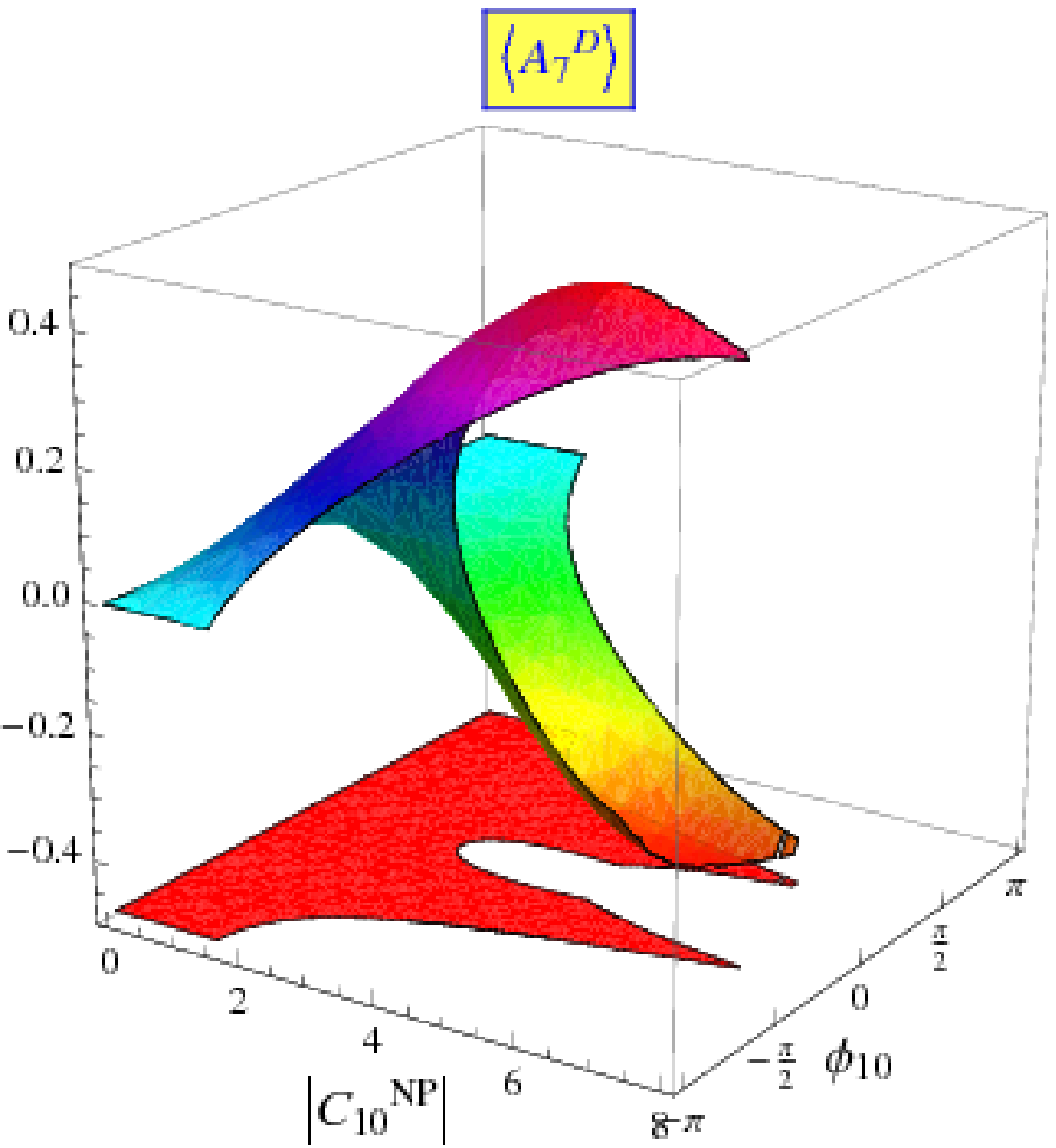, height=6cm, angle=0} &
 \epsfig{figure=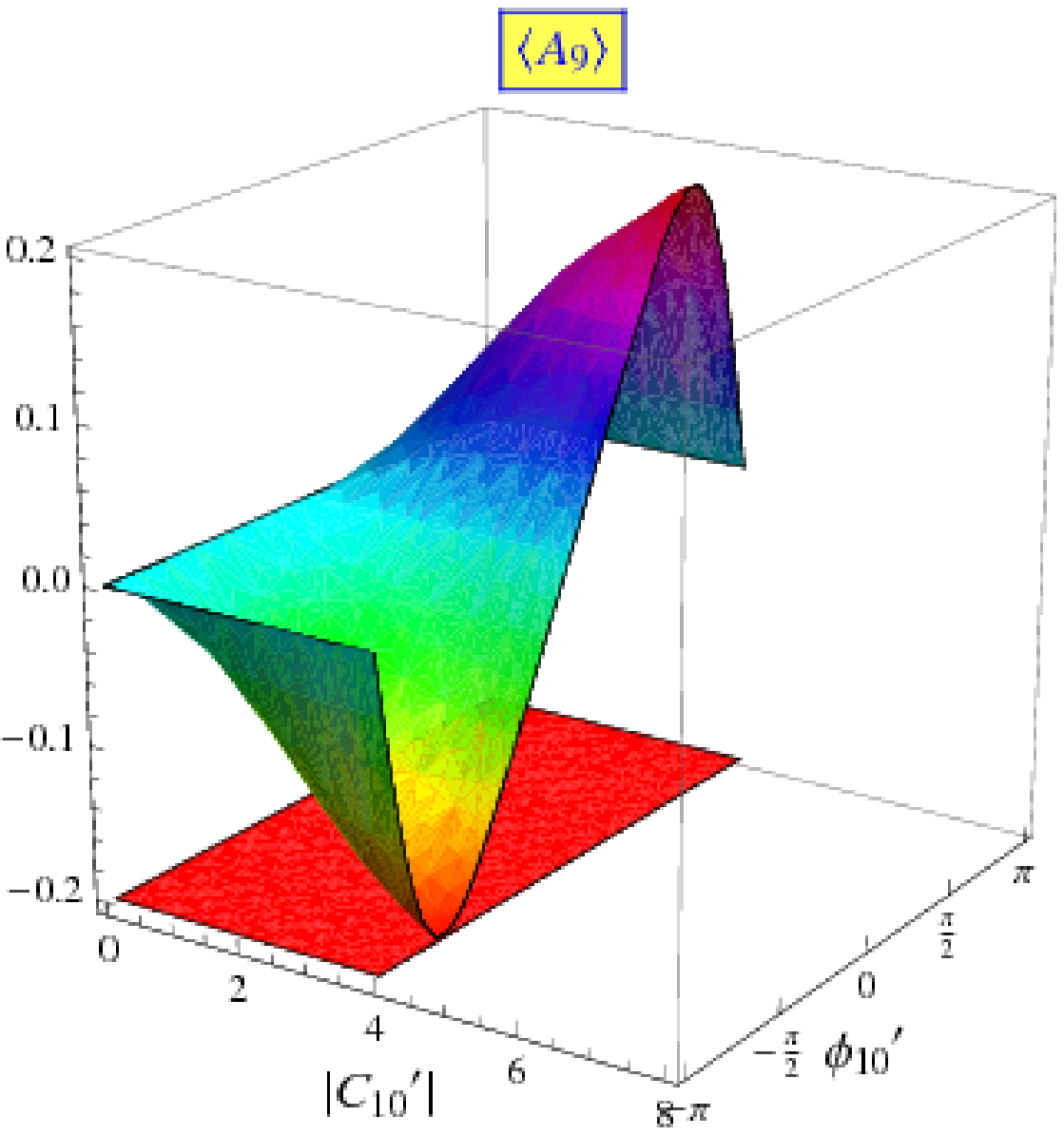, height=6cm, angle=0}\\
       \epsfig{figure=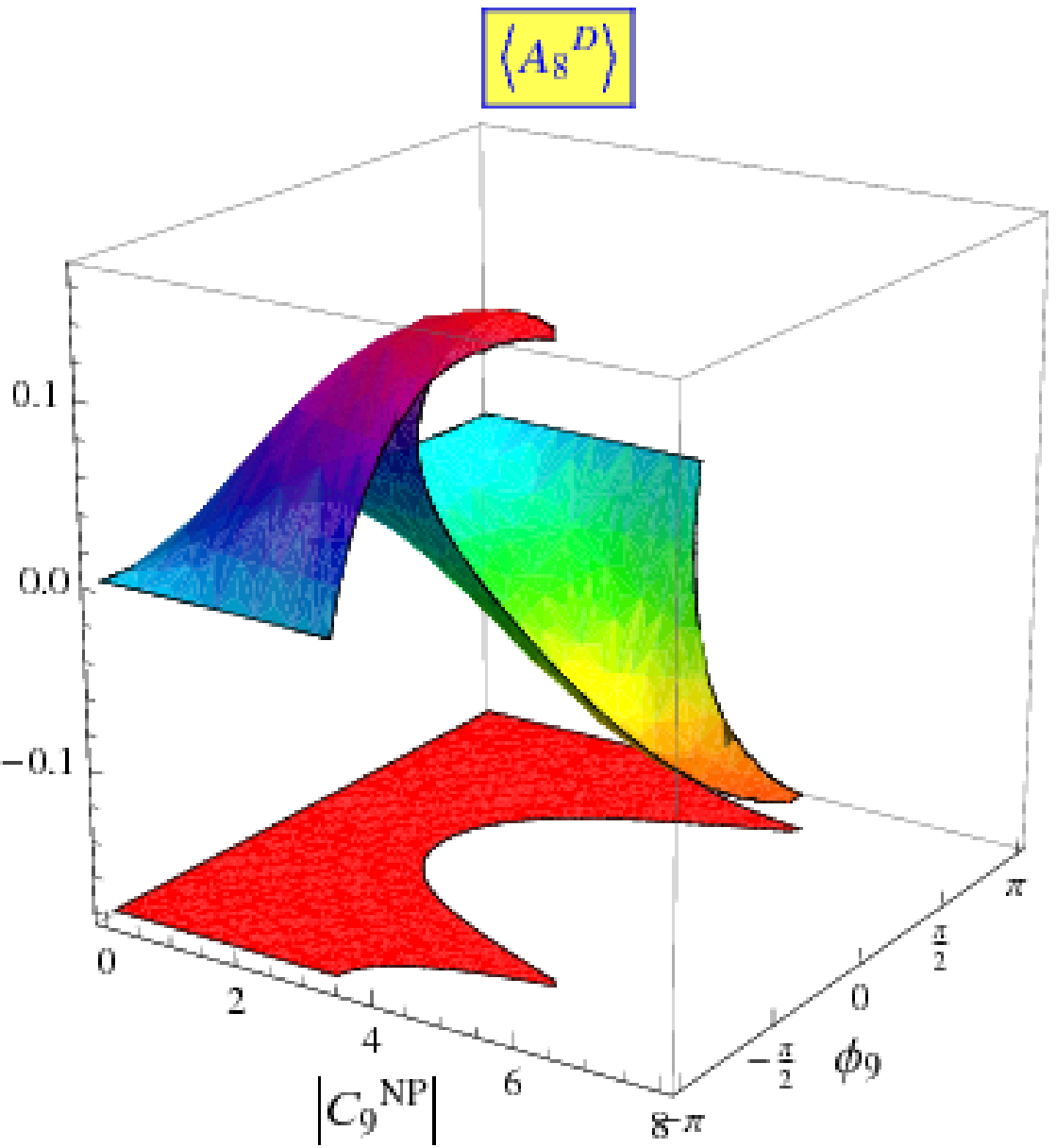, height=6cm, angle=0} &
   \epsfig{figure=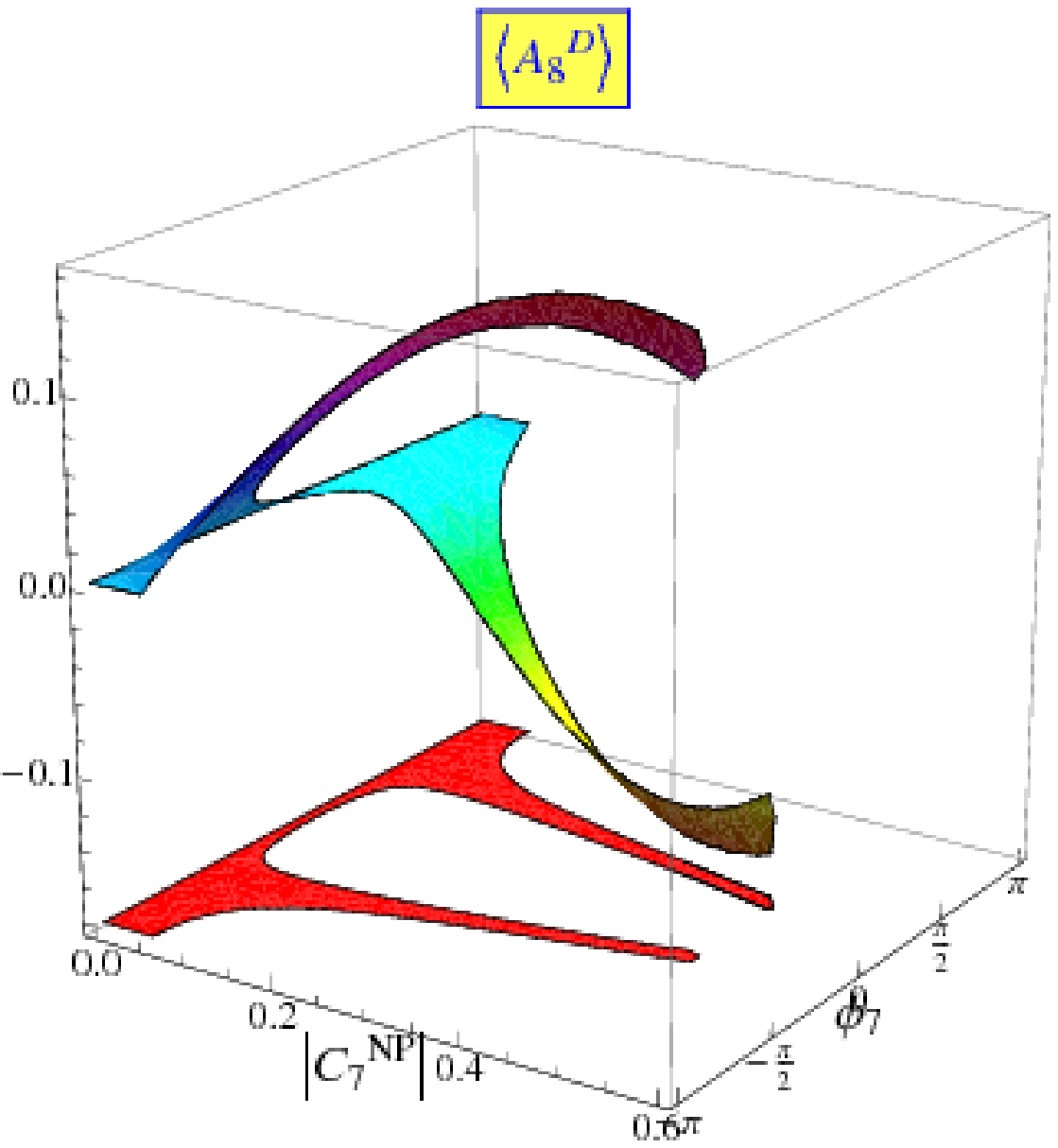, height=6cm, angle=0}
      \end{tabular}
 \end{center} \caption{ 
\label{fig:CPas7:NP}
The dependence of the integrated T-odd CP asymmetries 
$\langle A_{7,8,9}^{(D)} \rangle$
for $(q^2_{\rm min}, q^2_{\rm max}) = (1, 6) \GeV^2$ on NP Wilson coefficients 
after applying the experimental constraints. In each plot 
all other NP Wilson coefficients have been set to zero.}
 \end{figure}

We work out the CP asymmetries $\langle A_{i}^{(D)} \rangle$ with NP by taking 
into account the experimental constraints given in \reftab{tab:exp:data}.
We consider scenarios with generic NP, that is, when all six NP Wilson 
coefficients are varied independently, and when varying only one 
coefficient at a time. The asymmetries are integrated over low dilepton 
masses, $q^2 \in [1,6] \GeV^2$. Theoretical input 
parameters used are fixed at their central values.

In \reftab{tab:CPas:NP:res} we show the allowed ranges of the CP asymmetries 
in various NP scenarios. Numerically we find that the 
CP asymmetries can deviate significantly from their SM values, which
are doubly Cabibbo-suppressed and below the percent level. As anticipated, 
the T-even CP asymmetries can be enhanced by one order of magnitude up to 
$\lesssim 10\%$. The T-odd CP asymmetries $A_{7,8,9}^{(D)}$ can receive 
even stronger NP enhancements, up to order one.

There is also some residual dependence in the $A_i^{(D)}$ 
on all NP Wilson coefficients from the normalization to the CP averaged 
decay rate. Hence, even though the
numerator of some CP asymmetries is independent of a particular 
Wilson coefficient,
the asymmetries can be modified from their respective SM 
values given in \reftab{tab:CPas:SM:res}. These small effects are included in
\reftab{tab:CPas:NP:res} whenever they are distinguishable from the SM at 
$1 \sigma$, otherwise called SM-like.

Also the purely leptonic decay 
$\bar B_s \to \bar \mu \mu$ has strong sensitivity to
NP contributions in ${\cal{O}}_{10}$ and  ${\cal{O}}_{10}'$ since
$\BR(\Bstomm) \propto |C_{10} -C_{10}'|^2$, see, e.g., \cite{Buchalla:2000sk}. 
We find a possible enhancement of $\BR(\Bstomm)$ up to 
almost an order of magnitude in NP scenarios with these coefficients modified,
see \reftab{tab:CPas:NP:res}. The largest branching ratio, obtained 
with generic NP, is still a factor of two below the current experimental 
upper bound given in \reftab{tab:exp:data}. Furthermore, 
$\BR(\Bstomm)$ can be suppressed with respect to the SM by cancellations 
between $C_{10}$ and $C_{10}'$. A lower bound exists from data on the decays 
$B \to K^{(*)} \bar l l$, which are sensitive to 
$|C_{10} +  C_{10}'|$ \cite{Buchalla:2000sk}. However,
in models containing both $C_{10}^{\rm NP}$ and $C_{10}'$ 
only a very weak bound on  $\BR(\Bstomm)$ can be obtained.
We conclude that improved data on or a discovery of 
$\Bstomm$ decays will have a strong impact on this type of analysis.

In \reffig{fig:CPas7:NP} we show the dependence of the T-odd CP asymmetries 
$\langle A_{7,8,9}^{(D)} \rangle$ 
integrated over $(q^2_{\rm min}, q^2_{\rm max}) = (1, 6) \GeV^2$
on the NP Wilson coefficients as indicated. In the plots all other NP 
Wilson coefficients have been set to zero and the experimental FCNC 
constraints have been applied. The dependence of the asymmetries on the 
phases is very strong, making 
the CP asymmetries great probes of CP violation beyond the SM.

\begin{figure}
 \begin{center}
 \begin{tabular}{c}
   \epsfig{figure=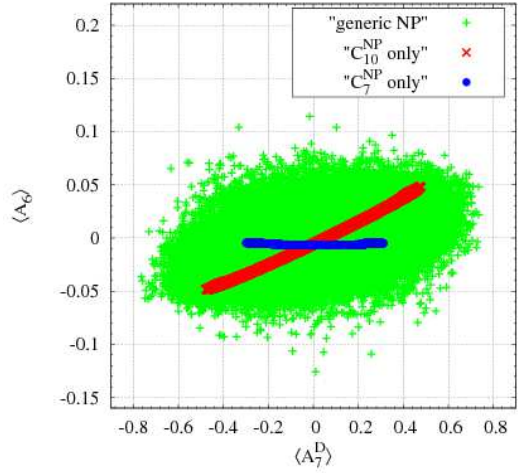,totalheight=8cm, angle=0}\\
   \\[-15mm]
   \end{tabular}
 \end{center}
 \caption{ \label{fig:sc:c10} Correlations between the 
CP asymmetries $\langle A_7^D \rangle $ 
and $\langle A_6 \rangle$ in different scenarios of NP:
generic NP, where all six coefficients are varied, and with NP in $C_{10}$
and $C_{7}$ only.}
\end{figure}

In \reffig{fig:sc:c10} $\langle A_6 \rangle $ is shown against
$\langle A_7^D \rangle$ for different NP scenarios. 
Both CP asymmetries are very sensitive
to the phase of  $C_{10}^{(')}$, and $A_7^D$ depends in addition
on the dipole coefficients $C_7^{(')}$. 
Correlations of this type can identify the nature of NP.

%
%--------+---------+---------+---------+---------+---------+---------+---------+
\section{Summary \label{sec:conclusions}}

We exploited the full angular analysis in exclusive semileptonic 
$\BtoKKpill$ and  $\bar B_s \to \phi (\to K^+ K^-) \bar l l$ decays
as a means of testing the SM and searching for new CP phases in 
$b \to s$ transitions. {}From the angular distributions in \refeq{eq:d4Gam} 
and \refeq{eq:d4barGam} seven CP asymmetries $A_{3 \ldots 9}^{(D)}$ 
in addition to the one in the decay rate can be accessed. We find that the 
SM predictions, valid for low dilepton masses, have rather large uncertainties 
$\sim 20 \%$ for $A_{4,5}^D$,
$\sim 50 \%$ for $A_{\rm CP}, A_6,A_{7,8}^D$ and
order one for $A_{3,9}$, but the tiny magnitude of the CP asymmetries 
$\lesssim 10^{-2}$ makes them all ideal
to search for a variety of different NP effects.
In particular, large NP effects are possible, which survive also 
the current experimental FCNC constraints.
We summarize here specific features of the asymmetries:
\begin{itemize}
\item[--] $A_7^D,A_8^D,A_9$ are T-odd and can be order one with New Physics.
\item[--] $A_5^D,A_6, A_8^D,A_9$ are CP-odd and can be obtained without
tagging from $d \Gamma + d \bar \Gamma$.
\item[--] $A_3,A_9$ are very sensitive to right-handed currents.
\item[--] $A_3,A_9, (A_6)$ can be extracted from a single-differential
distribution in $\phi (\theta_l)$.
\item[--] $A_7^D$ is very sensitive to the phase of the $Z$-penguins 
$\sim C_{10}^{(')}$.
\end{itemize}

Due to the CP-odd feature of some of the asymmetries and the width difference
in the $B_s$-system, further CP asymmetries $A_i^{(D) mix}$ can be extracted 
from untagged, time-integrated 
$\bar B_s, B_s \to \phi (\to K^+ K^-) \bar l l$ decays.
In the presence of NP with large CP phases we expect 
order one (order ten percent) 
CP asymmetries related to the T-odd (T-even) asymmetries
$A_{8,9}^{(D) mix}$ ($A_{5,6}^{(D) mix}$).
The difference between the asymmetries originating from
$B_d$- and $B_s$-decays is dominated by the difference in mixing.

A comparison of the $B_d$ and $B_s$ asymmetries hence probes the width 
difference and the mixing phase. One possibility is that the 
$A_i^{(D) mix}$ are order 10 percent, whereas $A_i^{(D)}$ are negligible, 
indicating that there is beyond-the-SM CP violation in $\Delta B =2$ 
processes only. Note that a measurement of the  $A_i^{(D)mix}$ is probably 
the only easy way to study CP violation in
semileptonic rare $b \to s$ transitions with $B_s$-mesons.

We conclude that the CP asymmetries from the angular analysis map out 
precisely the CP properties of several Wilson coefficients. 
The minimal, that is, CKM description of CP violation can be disproved.
This study can be extended to include also
lepton flavor dependent effects along the lines of \cite{Bobeth:2007dw}.

The prospects for studying rare dimuon modes at the LHC are promising: 
For an integrated luminosity of $2 fb^{-1}$, i.e., after one nominal year of 
data taking, a few thousand $B_d \to K^{*0} \mu^+ \mu^-$ events are expected
at LHCb, allowing a measurement of the branching ratio and its CP asymmetry
at the percent level \cite{lopez2005}. The CP asymmetries proposed here
require further information on angular distributions, thus higher statistics.
A dedicated sensitivity study, also taking into account suitable cuts, 
would be desireable.

%--------+---------+---------+---------+---------+---------+---------+---------+
\acknowledgments
 
G.H.~is happy to thank Frank Kr\"uger for early discussions on $A_7^D$
and Robert Fleischer for careful reading of the manuscript. 
C.B.~would like to thank Sebastian J\"ager for helpful explanations.
The work by C.B.~and G.H.~is supported in part by the Bundesministerium 
f\"ur Bildung und Forschung, Berlin-Bonn.  
G.P.~is supported by a grant from the
G.I.F., the German-Israeli-Foundation for Scientific Research and Development.
G.H.~is grateful to the organizers and participants of the
{\it Physics of the Large Hadron Collider}-workshop and
acknowledges the hospitality and great research
environment provided by the Kavli Institute for Theoretical 
Physics in Santa Barbara during the final phase of this work.

%
%
%--------+---------+---------+---------+---------+---------+---------+---------+
\appendix

%
%
%--------+---------+---------+---------+---------+---------+---------+---------+
\section{Angular Coefficients $J_i^{(a)}$ \label{app:Ii}}

Here the functions $J_i^{(a)}$ in the angular distribution \refeq{eq:d4Gam}
are given in terms of the transversity amplitudes $A_{\perp,\parallel,0,t}$
\cite{Kruger:2005ep}:
\begin{align}\label{I}
  J_1^s & = \frac{3}{4} \bigg\{ \frac{(2+\beta_l^2)}{4} \left[|\apeL|^2 + |\apaL|^2 + (L\to R) \right] 
            + \frac{4 m_l^2}{q^2} \re\left(\apeL^{}\apeR^* + \apaL^{}\apaR^*\right) \bigg\}, 
\\
  J_1^c & = \frac{3}{4} \bigg\{ |\azeL|^2 +|\azeR|^2  + \frac{4m_l^2}{q^2} 
               \left[|A_t|^2 + 2\re(\azeL^{}\azeR^*) \right]\bigg\},
\\
  J_2^s & = \frac{3 \beta_l^2}{16}\bigg[ |\apeL|^2+ |\apaL|^2 + (L\to R)\bigg],
\\
  J_2^c & = - \frac{3\beta_l^2}{4}\bigg[|\azeL|^2 + (L\to R)\bigg],
\\
  J_3 & = \frac{3}{8}\beta_l^2\bigg[ |\apeL|^2 - |\apaL|^2  + (L\to R)\bigg],
\\
  J_4 & = \frac{3}{4\sqrt{2}}\beta_l^2\bigg[\re (\azeL^{}\apaL^*) + (L\to R)\bigg],
\\
  J_5 & = \frac{3\sqrt{2}}{4}\beta_l\bigg[\re(\azeL^{}\apeL^*) - (L\to R)\bigg],
\\
  J_6 & = \frac{3}{2}\beta_l\bigg[\re (\apaL^{}\apeL^*) - (L\to R)\bigg],
\\
  J_7 & = \frac{3\sqrt{2}}{4} \beta_l \bigg[\im (\azeL^{}\apaL^*) - (L\to R)\bigg],
\\
  J_8 & = \frac{3}{4\sqrt{2}}\beta_l^2\bigg[\im(\azeL^{}\apeL^*) + (L\to R)\bigg],
\\
  J_9 & = \frac{3}{4}\beta_l^2\bigg[\im (\apaL^{*}\apeL) + (L\to R)\bigg],
\end{align}
where
\begin{equation}
\beta_l  = \sqrt{1 - \frac{4 m_l^2}{q^2}} ,
\end{equation}
and the
transversity amplitudes in QCDF can be seen in  \refeq{eq:trans:amp}.

%
%
%--------+---------+---------+---------+---------+---------+---------+---------+
\section{The effective Hamiltonian \label{sec:eff:Ham}}

We use the $\Delta B = 1$ effective Hamiltonian for $b \to s$ transitions,
e.g., \cite{Bobeth:1999mk, Beneke:2004dp}
\begin{align}
  \label{eq:Heff}
  {\cal{H}}_{\rm eff} & = -\frac{4 G_F}{\sqrt{2}} V_{tb}^{} V_{ts}^\ast
     \left( {\cal{H}}_{\rm eff}^{(t)} + \hat{\lambda}_u {\cal{H}}_{\rm eff}^{(u)} \right), & 
    \hat{\lambda}_u & = V_{ub}^{} V_{us}^\ast/V_{tb}^{} V_{ts}^\ast ,
\end{align}
where $V_{ij}$ denote CKM matrix elements and we used unitarity to write the
basis as
\begin{align}
  \label{eq:Heff:parts}
  {\cal{H}}_{\rm eff}^{(t)} & = C_1 \Op_1^c + C_2 \Op_2^c + \sum_{i = 3}^{10} C_i \Op_i, &  
  {\cal{H}}_{\rm eff}^{(u)} & = C_1 (\Op_1^c - \Op_1^u) + C_2 (\Op_2^c - \Op_2^u).
\end{align}
Here, the $\Op_{1,2}^{u,c}$ denote current-current operators whereas 
$\Op_i$ for $i=3,4,5,6$ are QCD-penguin operators, defined as in 
\cite{Chetyrkin:1996vx}. We further take into account the following
dipole and semileptonic operators
\begin{align}
  \Op_7 & = 
    \frac{e}{(4 \pi)^2} \overline m_b [\bar{s} \sigma^{\mu\nu} P_R b] F_{\mu\nu}, &
  \Op'_7 & = 
    \frac{e}{(4 \pi)^2} \overline m_b [\bar{s} \sigma^{\mu\nu} P_L b] F_{\mu\nu}, \nn
  \Op_9 & = 
    \frac{e^2}{(4 \pi)^2} [\bar{s} \gamma_\mu P_L b][\bar{l} \gamma^\mu l], &
  \Op'_9 & = 
    \frac{e^2}{(4 \pi)^2} [\bar{s} \gamma_\mu P_R b][\bar{l} \gamma^\mu l], \nn
  \Op_{10} & = 
    \frac{e^2}{(4 \pi)^2} [\bar{s} \gamma_\mu P_L b][\bar{l} \gamma^\mu \gamma_5 l],  &
  \Op'_{10} & = 
    \frac{e^2}{(4 \pi)^2} [\bar{s} \gamma_\mu P_R b][\bar{l} \gamma^\mu \gamma_5 l],
  \label{eq:basis}
\end{align}
where $P_{R/L} =(1 \pm \gamma_5)/2$ denote chiral projectors and $\overline
m_b(\mu_b)$ is the $\MSbar$ $b$-quark mass at the scale $\mu_b$. Since in
the SM $C'_i \sim m_s/m_b C_i$, the chirality flipped operators $\Op'_{7,9,10}$
can only compete with $\Op_{7,9,10}$ in models beyond the SM.
The Wilson coefficients are decomposed into their SM and NP parts as
$C_i = C_i^{\rm SM} + C_i^{\rm NP}$ and $C_i'=C_i'^{\rm NP}$ for $i = 7,9,10$.

%
%
%--------+---------+---------+---------+---------+---------+---------+---------+
\section{Transversity Amplitudes at NLO \label{sec:transamp}}

Starting from the $K^*$ transversity amplitudes $A_i(q^2)$,
$i = \{\perp, \parallel, 0, t\}$ in naive factorization (see, e.g.,
\cite{Kruger:2005ep}), the NLO $\alS$-corrections at large recoil
using QCDF \cite{Beneke:2001at,Beneke:2004dp} can be taken into
account by the replacements \cite{Kruger:2005ep, Lunghi:2006hc}
\begin{align}
  \label{eq:Wil:to:calT}
  (C_7^{\rm eff} + C_7^{' }) T_i(q^2) & \to {\cal T}^+_i, &
  (C_7^{\rm eff} - C_7^{' }) T_i(q^2) & \to {\cal T}^-_i, &
  C_9^{\rm eff}(q^2) & \to C_9 ,
\end{align}
where 
\begin{align}
  {\cal T}_1^\pm & = {\cal T}_\perp^\pm, &
  {\cal T}_2^- & = \frac{2 E}{M_B} {\cal T}_\perp^-, &
  {\cal T}_3^- & = {\cal T}_\perp^- + {\cal T}_\parallel^- .
\end{align}
The functions ${\cal T}_{\perp,\parallel}^-$  can be obtained from the 
${\cal T}_{\perp,\parallel}$  given in \cite{Beneke:2001at, Beneke:2004dp} by 
substituting $C_7^{\rm eff}$ with $C_7^{\rm eff} - C_7^{' }$ whereas  
${\cal T}_{\perp}^+$ is obtained from ${\cal T}_{\perp}$ by replacing 
$C_7^{\rm eff}$ with $C_7^{\rm eff} +C_7^{' }$.

In \refeq{eq:Wil:to:calT}, the $T_i$, $i=1,2,3$ denote the QCD tensor form 
factors defined in \refapp{app:formf}. The effective electroweak Hamiltonian 
employed is given in \refapp{sec:eff:Ham}. The effective coefficients
$C_{7,8}^{\rm eff}$ and $C_9^{\rm eff}(q^2)$ have been introduced to absorb 
1-loop matrix elements of 4-quark operators \cite{Buras:1993xp}. Here, such 
contributions to ${\cal{O}}_9$ are contained in ${\cal T}_i^\pm$ together
with further corrections beyond naive factorization. We take $C_{7}^{\rm eff}$ 
and $C_{9,10}$ at NNLL in the SM at the scale $\mu_b$. In the NP scenarios 
discussed in this work, $C_7^{' \rm eff}$ equals $C'_7$.

In the framework of QCDF, the functions ${\cal T}_{\perp,\parallel}^\pm$ are 
known at NLO in $\alS$ for the SM operators and the corresponding chirality 
flipped operators, see \refeq{eq:basis}. The ${\cal T}_{\perp,\parallel}^\pm$ 
have the following CKM and QCD structure 
\begin{align}
  \label{eq:calT:up:top}
  {\cal T}^\pm_a & = {\cal T}_a^{\pm (t)} + \hat\lambda_u {\cal T}_a^{(u)}, 
\end{align}
\begin{align}
  \nonumber 
  {\cal T}_a^{\pm (t)} & = {\cal T}_a^{\pm (t),\rm LO}
                         + \frac{\alS}{4\pi} {\cal T}_a^{\pm (t),\rm NLO}, &
  {\cal T}_a^{(u)} & = {\cal T}_a^{ (u),\rm LO} + \frac{\alS}{4 \pi} {\cal T}_a^{(u),\rm NLO},
\end{align}
where $a = \perp, \parallel$. At LO in $\alpha_s$ (denoted by the superscript $(0)$)
they read 
\begin{align} \label{eq:calT:LO}
  {\cal T}^{\pm (t), \rm LO}_{\bot} & = \xi_\perp \left[ 
    C_7^{ {\rm eff} (0)} \pm C_7^{'(0)} + \frac{q^2}{2 m_b M_B} Y^{(0)} \right],  &
  {\cal T}^{(u), \rm LO}_{\bot} & =  \xi_\perp \frac{q^2}{2 m_b M_B} Y^{(u) (0)}, \\
  {\cal T}^{-(t), \rm LO}_{\|} & = - \xi_\parallel \left[ 
   C_7^{\rm eff (0)} - C_7^{'(0)} + \frac{M_B}{2 m_b} Y^{(0)} \right] + HS, &
  {\cal T}^{(u), \rm LO}_{\|} & =
     -\xi_\parallel \frac{M_B}{2 m_b} Y^{(u)(0)} + HS,
  \nonumber
\end{align} 
where $Y(q^2)$ and $Y^{(u)}(q^2)$
contain 1-loop contributions of four-quark operators $\sim \bar s b \bar q q$ 
with an imaginary part for $q^2 > 4 m_q^2$. Since the charm threshold
is at the very upper end - if not outside - the $q^2$-region 
where the $1/E$ expansion works and the lighter quarks induce either 
CKM suppressed or penguin contributions, the resulting strong phase is 
small. In \refeq{eq:calT:LO}, spectator effects are denoted by $HS$. 
At lowest order, these are in 
${\cal T}^{(u),\rm LO}_{\|}$ and ${\cal T}^{- (t), \rm LO}_{\|}$. 
The latter is suppressed by
penguin coefficients, whereas the former is non-zero only
for charged ${B}^\pm \to K^{\ast \pm} \bar l l$ decays (weak annihilation). 
At higher order in $\alS$, strong  phases are further generated
in ${\cal T}_a^{(i),\rm NLO}$ and from spectator interactions 
\cite{Beneke:2001at, Beneke:2004dp}, 
which have been included in our numerical analysis.
The form factors $\xi_\perp$ and $\xi_\parallel$ are discussed in 
\refapp{app:formf}.

The transversity amplitudes in the presence of NP Wilson coefficients
within QCDF and neglecting kinematical terms\footnote{ These formally
subleading terms in the $1/E$ expansion are included in the numerical
evaluation.} $M_{K^*}^2/M_B^2$ read as
\begin{align}
  A_{\perp}^{L,R} & = + \sqrt{2} N M_B (1 - \hat{s}) \Bigg\{
         \Big[(C_9^{} + C'_9) \mp (C_{10}^{} + C'_{10})\Big] \xi_\perp
   + \frac{2 \hat{m}_b}{\hat{s}} {\cal T}_\perp^+ \Bigg\},
\nonumber \\
  A_{\parallel}^{L,R} & = - \sqrt{2} N M_B (1 - \hat{s}) \Bigg\{
         \Big[(C_9^{} - C'_9) \mp (C_{10}^{} - C'_{10})\Big] \xi_\perp
   + \frac{2 \hat{m}_b}{\hat{s}} {\cal T}_\perp^- \Bigg\},
\nonumber \\
  A_{0}^{L,R} & = - \frac{N M^2_B (1 - \hat{s})^2}{2 M_{K^*}
\sqrt{\hat{s}}} \Bigg\{
     \Big[(C_9^{} - C'_9) \mp (C_{10}^{} - C'_{10})\Big] \xi_\parallel
   - 2 \hat{m}_b {\cal T}_\parallel^- \Bigg\},
\nonumber \\
  A_{t} & = \frac{N M^2_B (1 - \hat{s})^2}{M_{K^*} \sqrt{\hat{s}}}
     (C_{10}^{} - C'_{10}) \frac{\xi_\parallel}{\Delta_\parallel} ,
  \label{eq:trans:amp}
\end{align}
where
\begin{align}
  \hat{s} & = \frac{q^2}{M_B^2}, &
  \hat{m}_b & = \frac{m_b}{M_B}, &
  N & = \Bigg[ \frac{\GF^2 \alE^2}{3\cdot 2^{10}\, \pi^5 M_B} 
     |V_{tb}^{} V_{ts}^\ast|^2\, \hat{s} \, \sqrt{\lambda} \, \beta_l \Bigg]^{1/2}
\end{align}
and $\lambda = M_B^4 + M_{K^*}^4 + q^4 - 2 (M_B^2 M_{K^*}^2 + M_B^2 q^2 + M_{K^*}^2 q^2)$.
Note that $A_t$ contributes only for $m_l \neq 0$ and contains $\Delta_\parallel$,
see \cite{Beneke:2001at}, which represents form factor symmetry breaking QCD corrections. 
Note that helicity conservation dictates $A_{\perp}^{L,R} = -A_{\parallel}^{L,R}$
for  $C_i'=0$ up to $1/E$ corrections \cite{Burdman:2000ku}.

The dilepton spectrum defined in \refeq{eq:rate} can be written 
in terms of the transversity amplitudes \refeq{eq:trans:amp} as
\begin{equation}
  \frac{d\Gamma}{dq^2} =
  |A_\perp^L|^2 + |A_\parallel^L|^2 + |A_0^L|^2
      + (L \to R)  + \order{m_l^2/q^2}
   \label{eq:dGdq:ff}
\end{equation}
up to contributions suppressed by the lepton mass. 
The dependence on the form factors stems from $\xi_\perp$ for 
$A_\perp^{L,R} , A_\parallel^{L,R}$ and $\xi_\parallel$ for $A_0^{L,R}$. 
Since the longitudinal amplitudes $A_0^{L,R}$ are enhanced by  $M_B/M_{K^*}$, 
see \refeq{eq:trans:amp}, they imply a
stronger dependence of $d\Gamma/dq^2$ on $\xi_\parallel$ than on $\xi_\perp$.
Quantitatively, we find in the SM that for the cuts 
$(q^2_{\rm min}, q^2_{\rm max}) = (1, 6) \GeV^2$  
and $(1, 7) \GeV^2$ the contribution from the longitudinal $K^*$ to the total
decay rate,
\begin{equation}
  {\cal F}_L = 
    \frac{\langle |A_0^L|^2 + |A_0^R|^2 \rangle }{\langle d \Gamma/d q^2 \rangle} ,
  \label{eq:Fl}
\end{equation}
is $0.73^{+0.08}_{-0.10}$ and $0.72^{+0.08}_{-0.11}$, respectively.

%
%
%--------+---------+---------+---------+---------+---------+---------+---------+
\section{CP Asymmetries and $A_{\rm FB}$ beyond the SM \label{app:CP:asy}}

Here we give analytical expressions of the CP asymmetries defined in
\refeq{eq:diff:CPasy:1} and \refeq{eq:diff:CPasy:2} including contributions
from NP operators \refeq{eq:basis}. The asymmetries
have been obtained from the transversity amplitudes in QCDF, 
see \refeq{eq:trans:amp}, valid in the low dilepton mass region.
The coefficients $C_7^{\rm eff} = C_7^{\rm eff, SM} + C_7^{\rm NP}$ and 
$C_7'$ are taken into account 
by ${\cal T}_{\perp,\parallel}^\pm$. Except for $A_{\rm CP}$, the 
CP asymmetries are given with their full lepton mass dependence which 
is confined to powers of $\beta_l$. Neglecting kinematical factors 
$M_{K^*}^2/M_B^2$, the CP asymmetries as a function of the dilepton mass 
can be written as
\begin{align}
  A_{\rm CP} & = {\cal A} \frac{8 \hat{m}_b}{3 \hat{s}} \re \bigg\{
    \frac{\xi_\parallel^2}{\xi_\perp^2} \frac{M_B^2}{M_{K^*}^2}
      \frac{(1 - \hat{s})^2}{8} \left[ 
       \hat{m}_b \frac{|{\cal T}_\parallel^-|^2}{\xi_\parallel^2} -
       \frac{{\cal T}_\parallel^-}{\xi_\parallel} (C_9^{} - C'_9)^\ast \right]
   + \frac{\hat{m}_b}{\hat{s}} \frac{|{\cal T}_\perp^+|^2 + |{\cal T}_\perp^-|^2}{\xi_\perp^2}    
\nonumber \\
  & \hspace{1cm}
  + \frac{{\cal T}_\perp^+ - {\cal T}_\perp^-}{\xi_\perp} C_9^{*}
  + \frac{{\cal T}_\perp^+ + {\cal T}_\perp^-}{\xi_\perp} C_9'^{*} 
    - (\delta_W \to -\delta_W)  \bigg\} + \order{m_l^2/q^2},
\\[2ex]
  A_3 & = {\cal A} \frac{2 \hat{m}_b \beta_l}{\hat{s}} \re \bigg\{
    \frac{\hat{m}_b}{\hat{s}} \frac{|{\cal T}_\perp^+|^2 - |{\cal T}_\perp^-|^2}{\xi_\perp^2} 
  + \frac{{\cal T}_\perp^+ - {\cal T}_\perp^-}{\xi_\perp} C_9^{*}
  + \frac{{\cal T}_\perp^+ + {\cal T}_\perp^-}{\xi_\perp} C_9'^{*}    
  - (\delta_W \to -\delta_W)  \bigg\},
\\[2ex] 
  A_4^D & = - {\cal A}^D \frac{\hat{m}_b \beta_l}{2 \hat{s}} \re \bigg\{
    \left( \frac{{\cal T}_\perp^-}{\xi_\perp} 
     - \hat{s} \frac{{\cal T}_\parallel^-}{\xi_\parallel} \right) (C_9^{} - C'_9)^*  
  - 2 \hat{m}_b \frac{{\cal T}_\perp^- ({\cal T}_\parallel^-)^\ast}{\xi_\perp \xi_\parallel} 
  - (\delta_W \to - \delta_W)  \bigg\},
\\[2ex]
  A_5^D & = -{\cal A}^D  \frac{\hat{m}_b}{\hat{s}} \re \bigg\{
    \left( \frac{{\cal T}_\perp^-}{\xi_\perp} 
     - \hat{s} \frac{{\cal T}_\parallel^-}{\xi_\parallel} \right) C^{}_{10}
  - \left( \frac{{\cal T}_\perp^-}{\xi_\perp} 
     + \hat{s} \frac{{\cal T}_\parallel^-}{\xi_\parallel} \right) C_{10}'^{*} 
   - (\delta_W \to - \delta_W)  \bigg\},
\\[2ex] 
  A_6 & = {\cal A} \frac{4 \hat{m}_b}{\hat{s}} \re \bigg\{
    \frac{{\cal T}_\perp^+ + {\cal T}_\perp^-}{\xi_\perp} C_{10}^{*}
  - \frac{{\cal T}_\perp^+ - {\cal T}_\perp^-}{\xi_\perp} C_{10}'^{*} 
  - (\delta_W \to - \delta_W) \bigg\},
\\[2ex]
  A_7^D & = {\cal A}^D \frac{\hat{m}_b}{\hat{s}} \im \bigg\{
     (C^{}_{10} - C'_{10})
         \left( \frac{{\cal T}_\perp^-}{\xi_\perp} 
              + \hat{s} \frac{{\cal T}_\parallel^-}{\xi_\parallel} \right)^\ast 
  - (\delta_W \to -\delta_W) \bigg\}, 
\\[2ex]
  A_8^D & = {\cal A}^D \frac{\beta_l}{2} \, \im \bigg\{
     \frac{2 \hat{m}_b^2}{\hat{s}} 
       \frac{{\cal T}_\perp^+ ({\cal T}_\parallel^-)^\ast}{\xi_\perp \xi_\parallel}
   - \frac{\hat{m}_b}{\hat{s}} \bigg[
     \left( \frac{{\cal T}_\perp^+}{\xi_\perp} + \hat{s} \frac{{\cal T}_\parallel^-}{\xi_\parallel} \right) C_9^\ast  
   - \left( \frac{{\cal T}_\perp^+}{\xi_\perp} - \hat{s} \frac{{\cal T}_\parallel^-}{\xi_\parallel} \right) C_9'^{\ast}
        \bigg]
\nonumber \\ 
  &      \hspace{2.5cm}
        + C_9^{} C_9'^{*} + C_{10}^{} C_{10}'^{*} - (\delta_W \to -\delta_W)  \bigg\},
\\[2ex]
  A_9 & = - {\cal A}\, 2 \beta_l \im \bigg\{
    \frac{2 \hat{m}_b^2 }{\hat{s}^2} \frac{{\cal T}_\perp^+ ({\cal T}_\perp^-)^\ast}{\xi_\perp^2}  
  + \frac{\hat{m}_b}{\hat{s}} \bigg[
       \frac{{\cal T}_\perp^+ - {\cal T}_\perp^-}{\xi_\perp} C_9^{*}
     - \frac{{\cal T}_\perp^+ + {\cal T}_\perp^-}{\xi_\perp} C_9'^{*} \bigg] 
\nonumber \\ 
  &      \hspace{2.5cm} - C_9^{} C_9'^{*} - C_{10}^{} C_{10}'^{*} - (\delta_W \to -\delta_W)  \bigg\},
\end{align}
where $(\delta_W \to -\delta_W)$ is short hand notation for conjugating 
all weak phases. Furthermore,
\begin{align}
  {\cal A}   & = \frac{G_F^2\, \alE^2}{3^2 \cdot2^{6}\, \pi^5} |V_{tb}  ^{}V_{ts}^*|^2
    \frac{M_B^3 \beta_l^2 \hat{s} (1 - \hat{s})^3 \xi_{\bot}^2 }{N_\Gamma} , 
\nonumber \\    
  {\cal A}^D & = \frac{G_F^2\, \alE^2}{3^2 \cdot2^{6}\, \pi^5} |V_{tb}  ^{}V_{ts}^*|^2
    \frac{M_B^4 \beta_l^2 \sqrt{\hat{s}} (1 - \hat{s})^4 \xi_{\|}\xi_{\bot} }{M_{K^*} N_\Gamma} ,
\end{align}
where $N_\Gamma$  is defined in \refeq{eq:diff:CPasy:1}.

At lowest order in $\alS$, the expressions for the above CP asymmetries
simplify by
\begin{align}
   \frac{{\cal T}_\perp^{+, \rm LO} - {\cal T}_\perp^{-, \rm LO}}{\xi_\perp}
   & = 2 C_7^{' (0)},
\\[2ex]
   \frac{{\cal T}_\perp^{+, \rm LO} + {\cal T}_\perp^{-, \rm LO}}{\xi_\perp}
   & = 2 C_7^{\rm eff (0) } + \frac{\hat{s}}{ \hat m_b} ( Y^{(0)}
+ \hat\lambda_u Y^{(u) (0)}),
\\[2ex]
  \frac{{\cal T}_\perp^{\pm, \rm LO}}{\xi_\perp} +
    \hat{s} \frac{{\cal T}_\parallel^{-, \rm LO}}{\xi_\parallel} & =
   \left\{ \begin{array}{c} (1 - \hat{s}) C_7^{\rm eff (0)} + (1 + \hat{s})
C_7^{' (0)} \\[2ex]
                            (1 - \hat{s}) (C_7^{\rm eff (0) } - C_7^{' (0)})
\end{array}
  \right.  .
\end{align}
Note that in the SM, or more general, in any model without
right-handed contributions to the electromagnetic dipole operator, 
${\cal T}_\perp^+ ={\cal T}_\perp^-$, see \refapp{sec:transamp}.
 
The lepton forward-backward asymmetry in QCDF is written as
\begin{align}
  A_{\rm FB} & = \frac{12 \beta_l N^2 M_B^2 (1 - \hat{s})^2 \xi_\perp^2}{d\Gamma/dq^2}
  \\ 
  &      \hspace{1cm} \times
   \re \bigg\{ 
   \left[ C_9^{} + \frac{\hat{m}_b}{\hat{s}} \frac{({\cal T}^+_\perp + {\cal T}^-_\perp)}{\xi_\perp} \right] C_{10}^\ast 
 - \left[ C'_9   + \frac{\hat{m}_b}{\hat{s}} \frac{({\cal T}^+_\perp - {\cal T}^-_\perp)}{\xi_\perp} \right] C_{10}^{'\ast}  
  \bigg\}.
  \nonumber
\end{align}

%
%
%--------+---------+---------+---------+---------+---------+---------+---------+
\section{$\BtoKast$ Form Factors at Large Recoil \label{app:formf}}

The $\BtoKast$ matrix element can be parametrized in terms of seven 
$q^2$-dependent QCD form factors $V,A_{0,1,2}$ and $T_{1,2,3}$ as
\begin{align}
  & \langle K^*(p_B-q) | \bar{s} \gamma_\mu (1-\gamma_5) b |B(p_B) \rangle =
   - 2 \epsilon_{\mu\nu\alpha\beta} \varepsilon^{*\nu} p_B^\alpha q^\beta \frac{V}{M_B + M_{K^*}}
\\  
  & \hspace{0.5cm}
    - i \varepsilon_\mu^* (M_B + M_{K^*}) A_1
    + i (2 p_B - q)_\mu (\varepsilon^*\cdot q) \frac{A_2}{M_B + M_{K^*}} 
    + i q_\mu (\varepsilon^*\cdot q) \frac{2 M_{K^*}}{q^2} [A_3 - A_0],
\nonumber
\end{align}
\begin{align}
  & \langle K^*(p_B-q) | \bar{s} \sigma_{\mu\nu} q^\nu (1+\gamma_5) b | B(p_B) \rangle =
  -2 i\, \epsilon_{\mu\nu\alpha\beta} \varepsilon^{*\nu} p_B^\alpha q^\beta\, T_1
\\  
  & \hspace{0.5cm}  
  + [\varepsilon_\mu^* (M_B^2 - M_{K^*}^2) - (\varepsilon^*\cdot q) (2 p_B - q)_\mu]\, T_2 
  + (\varepsilon^*\cdot q) \bigg[q_\mu - \frac{q^2}{M_B^2 - M_{K^*}^2} (2 p_B - q)_\mu \bigg]\, T_3
\nonumber  
\end{align}
and
\begin{equation}
  A_3 = \frac{M_B + M_{K^*}}{2 M_{K^*}} A_1 - \frac{M_B - M_{K^*}}{2 M_{K^*}} A_2,
\end{equation}
where $\epsilon^{* \mu}$ denotes the polarization vector of the $K^*$ and
$p_B^\mu$ the four momentum of the $B$ meson.
The QCD form factors obey symmetry relations in the large recoil limit and can
be expressed at leading order in the $1/E$ expansion in terms of two universal 
form factors $\xi_\perp$ and $\xi_\parallel$ \cite{Charles:1998dr}.
Symmetry breaking corrections at order $\alS$ 
have been calculated using QCDF in Ref.~\cite{Beneke:2000wa}. 
We employ a factorization scheme within QCDF 
where the $\xi_{\perp, \parallel}$ are related to the $V, A_{1,2}$ 
as \cite{Beneke:2004dp}
\begin{align}
  \label{eq:xi:def}
  \xi_\perp & = \frac{M_B}{M_B + M_{K^*}} V, &
  \xi_\parallel & = \frac{M_B + M_{K^*}}{2 E} A_1 -
                    \frac{M_B - M_{K^*}}{M_B} A_2.
\end{align}

\begin{figure}
\begin{center}
 \begin{tabular}{c}
  \epsfig{figure=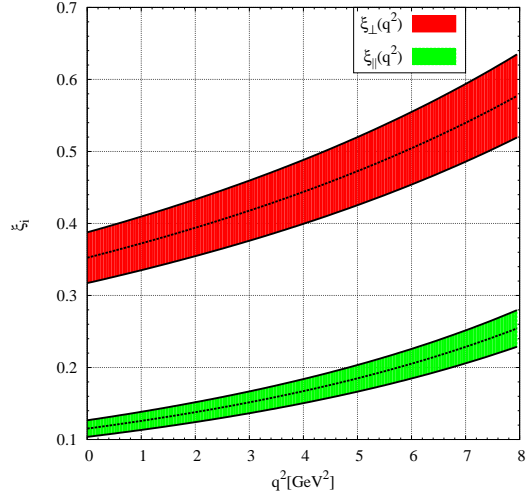,      height=8.0cm, angle=-0} 
\\[-10mm]
\end{tabular}
\end{center}
\caption{ \label{fig:xi:q2dep}
  The universal form factors $\xi_\perp$ and $\xi_\parallel$ in the low-$q^2$ 
  region and their uncertainty indicated by the bands. }
\end{figure}

For the  $q^2$ dependence of the universal form factors we adopt the findings
from light cone sum rule (LCSR) calculations \cite{Ball:2004rg}.
Here the $q^2$ dependence is parametrized as
\begin{align}
  V(q^2) & = \frac{r_1}{1 - q^2/m_R^2} + \frac{r_2}{1 - q^2/m_{fit}^2} ,
\\ 
  A_1(q^2) & = \frac{r_2}{1 - q^2/m_{fit}^2} ,
\\ 
  A_2(q^2) & = \frac{r_1}{1 - q^2/m_{fit}^2} + \frac{r_2}{(1 - q^2/m_{fit}^2)^2} ,
\end{align}
where the fit parameters $r_{1,2}, m^2_{R}$ and $m^2_{fit}$ are shown in
\reftab{tab:FF:fit}. Also given in this table are the values of the form
factors at $q^2=0$ and the corresponding parametric uncertainties within
the LCSR approach. We give the uncertainties independent of the Gegenbauer
moments $a^{\perp,\parallel}_{1,K^*}$ and the ones due to 
$a^{\perp,\parallel}_{1,K^*}$ separately. 
The relative uncertainty of the form factors $V(0), A_1(0)$ and $A_{2}(0)$ 
amounts to $8\%$, $10\%$ and $10\%$  without, and $11\%$, $12\%$ and $14\%$
after adding the $a_{1,K^*}$ -- see \reftab{tab:num:input} for the numerical
value -- induced uncertainty in quadrature, respectively. We use the total
relative uncertainty from maximal recoil as an estimate for the form factor
uncertainties for $q^2>0$. 
The form factors $\xi_{\perp, \parallel}$ defined via \refeq{eq:xi:def}
are shown as a function of $q^2$ in \reffig{fig:xi:q2dep}. Here the bands 
indicate the uncertainty in $\xi_\perp$ and $\xi_\parallel$ of $11\%$ and
$14\%$, respectively.

\begin{table}[hb]
\centering
\begin{tabular}{||c||cccc|ccc||}
\hline \hline
  $ $ & $r_1$ & $r_2$ & $m_R^2\,[\GeV^2]$ & $m_{fit}^2\,[\GeV^2]$ 
      & $F(0)$  & $\Delta_0 F(0)$ & $\Delta_{a_1} F(0)$ \\
\hline
  $V$   & $0.923$  & $-0.511$ & $5.32^2$ & $49.40$ & $0.411$ & $0.033$ & $0.44 \delta_{a_1}$
\\[0.5ex]
%\hline
  $A_1$ &  \mbox{} & $ 0.290$ &  \mbox{} & $40.38$ & $0.292$ & $0.028$ & $0.33 \delta_{a_1}$ 
\\[0.5ex]
%\hline
  $A_2$ & $-0.084$ &  $0.342$ &  \mbox{} & $52.00$ & $0.259$ & $0.027$ & $0.31 \delta_{a_1}$
\\[0.5ex]
\hline \hline
\end{tabular}
\caption{\label{tab:FF:fit} The parameters $r_{1,2},m^2_{R}$ and $m^2_{fit}$ 
  describing the $q^2$ dependence of the form factors 
  $V$ and $A_{1,2}$ in the LCSR
  approach  \cite{Ball:2004rg}. Also shown are the corresponding values of 
  the form factors at $q^2=0$, $F(0)$, their uncertainties
  independent of the Gegenbauer moment $a_{1,K^*}$,  $\Delta_0 F(0)$ and the 
  uncertainties induced by $a_{1,K^*}$ in terms of 
   $\delta_{a_1} = (a_{1,K^*}(1 \GeV) - 0.1)$, $\Delta_{a_1} F(0)$. }
\end{table}

%
%
%--------+---------+---------+---------+---------+---------+---------+---------+
\section{Model-independent CP Asymmetries beyond the SM
  \label{app:num:CP:asy}}

%\TABLE[bt]{
\begin{table}
\begin{tabular}{||c|c|rrrrrr||}
\hline \hline
$X_{\rm SM}$ &  &
$i=7$ & $i=7'$ & $i=9$ & $i=9'$ & $i=10$ & $i=10'$
\\
\hline
  \multirow{2}{*}{$\BR_{\rm SM}$} &
  $a_i$ & $2.634$ & $2.634$ & $0.035$ & $0.035$ & $0.035$ & $0.035$
\\
& $b_i$ & $-0.271$ & $-0.373$ & $0.162$ & $-0.179$ & $-0.288$ & $0.205$
\\
  \multirow{1}{*}{$ = 2.444 \cdot 10^{-7}$} &
  $c_i$ & $-0.156$ & $0.001$ & $-0.009$ & $-0.0002$ & $0$ & $0$
\\
\hline
  \multirow{2}{*}{$\overline{\BR}_{\rm SM}$} &
  $a_i$ & $2.656$ & $2.656$ & $0.036$ & $0.036$ & $0.035$ & $0.035$
\\
& $b_i$ & $-0.312$ & $-0.370$ & $0.158$ & $-0.178$ & $-0.290$ & $0.206$
\\
  \multirow{1}{*}{$ = 2.423 \cdot 10^{-7}$} &
  $c_i$ & $0.106$ & $0.003$ & $0.004$ & $0.002$ & $0$ & $0$
\\
\hline
  ${\rm Num} \aver{A_{\rm CP}}_{\rm SM}$ &
  $b_i$ & $4.469$ & $- 0.726$ & $0.587$ & $-0.345$ & $0$ & $0$
\\
  $= 2.068 \cdot 10^{-9}$ &
  $c_i$ & $-30.770$ & $- 0.275$ & $- 1.500$ & $-0.259$ & $0$ & $0$
\\
\hline
  ${\rm Num} \aver{A_3}_{\rm SM}$ &
  $b_i$ & $-0.077$ & $5.720$ & $-0.012$ & $0.378$ & $0$ & $0$
\\
  $ = 0^\dagger$ &
  $c_i$ & $0.542$ & $-47.174$ & $0.081$ & $-2.743$ & $0$ & $0$
\\
\hline
  ${\rm Num} \aver{A_4^D}_{\rm SM}$ &
  $b_i$ & $3.604$ & $-3.604$ & $0.536$ & $-0.536$ & 0 & 0
\\
  $ = -8.642 \cdot 10^{-10}$ &
  $c_i$ & $-1.435$ & $1.435$ & $-2.487$ & $2.487$ & 0 & 0
\\
\hline
  ${\rm Num} \aver{A_5^D}_{\rm SM}$ &
  $b_i$ & 0 & 0 & 0 & 0 & $-0.244$ & $0.068$
\\
  $ = 3.718 \cdot 10^{-9}$ &
  $c_i$ & 0 & 0 & 0 & 0 & $1.152$ & $-1.258$
\\
\hline
  ${\rm Num} \aver{A_6}_{\rm SM}$ &
  $b_i$ & 0 & 0 & 0 & 0 & $-0.244$ & $0.004$
\\
  $ = -3.117 \cdot 10^{-9}$ &
  $c_i$ & 0 & 0 & 0 & 0 & $1.774$ & $-0.026$
\\
\hline
  ${\rm Num} \aver{A_7^D}_{\rm SM}$ &
  $b_i$ & $0$ & $0$ & $0$ & $0$ & $-0.244$ & $0.244$
\\
  $ = -2.496 \cdot 10^{-9}$ &
  $c_i$ & $-247.248$ & $247.248$ & $0$ & $0$ & $23.019$ & $-23.019$
\\
\hline
  ${\rm Num} \aver{A_8^D}_{\rm SM}$ &
  $b_i$ & $-0.491$ & $-1.423$ & $0.176$ & $-0.288$ & $0$ & $0$
\\
  $ = 1.706 \cdot 10^{-9}$ &
  $c_i$ & $-189.333$ & $-170.364$ & $-16.524$ & $-7.160$ &  $0$ & $26.834$
\\
\hline
  ${\rm Num} \aver{A_9}_{\rm SM}$ &
  $b_i$ & $0$ & $-8.390$ & $0.007$ & $-0.491$ & $0$ & $0$
\\
  $ = 0^\dagger$ &
  $c_i$ & $-6.514$ & $225.487$ & $-0.568$ & $6.064$ & $0$ & $31.913$
\\ \hline \hline
\end{tabular}
\caption{ \label{tab:gen:abc} The SM predictions $X_{\rm SM}$ and the
corresponding coefficients
  $a_i$, $b_i$ and $c_i$ for $i = 7, 7', 9, 9', 10, 10'$. ${}^\dagger$For
${\rm Num} \aver{A_{3, 9}}$
  $X_{\rm SM}$ has been set to zero and the corresponding coefficients are
given in units
  of $10^{-9}$.}
\end{table}
%}

%\TABLE[bt]{
\begin{table}
\centering
\begin{tabular}{||c|rr||c|rrr||}
\hline \hline
$d_{ij}$ & $\BR$ & $\overline{\BR}$ &
$e_{ij}$ & ${\rm Num} \aver{A_7^D}$ & ${\rm Num} \aver{A_8^D}$ & ${\rm Num}
\aver{A_9}^\dagger$
\\
\hline
$7,7'$ & $-0.255$ & $-0.257$ &
$7,7'$ & $0$ & $200.542$ & $1801.269$
\\
$7,9$ & $0.394$ & $0.397$ &
$7,9$ & $0$ & $-43.413$ & $-1.547$
\\
$7,9'$ & $-0.107$ & $-0.108$ &
$7,9'$ & $0$ & $56.532$ & $105.869$
\\
$7,10$ & $0$ & $0$ &
$7,10$ & $60.420$ & $0$ & $0$
\\
$7,10'$ & $0$ & $0$ &
$7,10'$ & $-60.420$ & $0$ & $0$
\\
\hline
$7',9$ & $-0.107$ & $-0.108$ &
$7',9$ & $0$ & $-56.532$ & $-105.869$
\\
$7',9'$ & $0.394$ & $0.397$ &
$7',9'$ & $0$ & $43.413$ & $1.547$
\\
$7',10$ & $0$ & $0$ &
$7',10$ & $-60.420$ & $0$ & $0$
\\
$7',10'$ & $0$ & $0$ &
$7',10'$ & $60.420$ & $0$ & $0$
\\
\hline
$9,9'$ & $-0.050$ & $-0.050$ &
$9,9'$ & $0$ & $6.558$ & $7.799$
\\
\hline
$10,10'$ & $-0.050$ & $-0.050$ &
$10,10'$ & $0$ & $6.558$ & $7.799$
\\
\hline \hline
\end{tabular}
\caption{ \label{tab:gen:de} The coefficients $d_{ij}$ and $e_{ij}$ for
  $i,j = 7, 7', 9, 9', 10, 10'$ and $j>i$. ${}^\dagger$For ${\rm Num}
  \aver{A_9}$ $X_{\rm SM}$ has been set to zero and the corresponding
  coefficients are given in units of $10^{-9}$.}
\end{table}
%}

We give numerical formulae for the $q^2$-integrated quantities
$\BR = \tau_{B^0} \aver{d\Gamma/dq^2}$, $\overline{\BR} =
\tau_{B^0} \aver{d\bar\Gamma/dq^2}$ and ${\rm Num} \aver{A_i^{(D)}}$ for
$q^2 \in [1, 6] \GeV^2$ in terms of the NP Wilson coefficients $C_i^{\rm NP}$. 
Here,
${\rm Num} \aver{A_i^{(D)}}$ denotes the numerators of the CP asymmetries
multiplied by the $B$-meson lifetime such that the normalized 
CP asymmetries (see \refeq{eq:averageAi}) are obtained from
\begin{equation}
  \aver{A_i^{(D)}} = \frac{{\rm Num} \aver{A_i^{(D)}}}{\BR + \overline{\BR}}.
\end{equation}
The dependence of the branching ratios on the NP Wilson coefficients can be 
written as 
\begin{align}
  X & = X_{\rm SM} \Big[ 1 +
     \sum_i \big( a_i |C_i^{\rm NP}|^2 + b_i \re C_i^{\rm NP} + c_i 
\im C_i^{\rm NP} \big)
   + \sum_{j>i} d_{ij} \re( C_i^{\rm NP} C_j^{\rm NP *} ) \Big]   
~~\mbox{for}~\BR,\overline{\BR},
\end{align}
whereas the numerators of the T-odd CP asymmetries are parametrized as
\begin{align}
  X & = X_{\rm SM} \Big[ 1 + \sum_i \big( b_i \re C_i^{\rm NP} 
+ c_i \im C_i^{\rm NP} \big)
   + \sum_{j>i} e_{ij} \im( C_i^{\rm NP} C_j^{\rm NP *} ) \Big]   ~\mbox{for}~{\rm Num} \aver{A_{7,8}^D} .
\end{align}
The numerators of the T-even CP asymmetries read as
\begin{align}
  X & = X_{\rm SM} \Big[ 1 + \sum_i \big(b_i \re C_i^{\rm NP} + c_i \im C_i^{\rm NP}  \big)
\Big] ~~~~\mbox{for}~{\rm Num} \aver{A_{\rm CP,6}}, {\rm Num} \aver{A_{4,5}^D}.
\end{align}
Here, the summations are over $i,j = {7, 7', 9, 9', 10, 10'}$ and $X_{\rm SM}$
denotes the SM prediction of the corresponding quantity. Note that 
for ${\rm Num} \aver{A_{3,9}}$ we have 
set $X_{\rm SM}$ to zero, see \refsec{sec:SM:num}, and, hence, 
the corresponding formulae read as
\begin{align}
  X & = \sum_i \big(b_i \re C_i^{\rm NP} + c_i \im C_i^{\rm NP}  \big)  ~~~~\mbox{for}~{\rm Num} \aver{A_3}, \\
  X & = \sum_i \big( b_i \re C_i^{\rm NP} + c_i \im C_i^{\rm NP} \big)
   + \sum_{j>i} e_{ij} \im( C_i^{\rm NP} C_j^{\rm NP *} ) ~~~~\mbox{for}~{\rm Num} \aver{A_9}.
\end{align}
The SM predictions $X_{\rm SM}$
and the coefficients $a_i$, $b_i$, $c_i$ and $d_{ij}$, $e_{ij}$ are given in
\reftab{tab:gen:abc} and \reftab{tab:gen:de}, respectively. We assumed 
central values for all parameters.

%--------+---------+---------+---------+---------+---------+---------+---------+

\end{document}